\newif\ifTwoColumn%
\newif\ifSUBMIT%
\newif\ifCOMMENTS%
\newif\ifFIGs%
\newif\ifFIGoneColumn%
 \let\ifFIGs\iftrue
 \let\ifFIGs\iffalse
\let\ifFIGoneColumn\iftrue%
\documentclass[aip,reprint,jcp,amsmath,amssymb,superscriptaddress,floatfix]{revtex4-1}

\usepackage{graphicx}
\usepackage{amsmath}
\usepackage{xcolor}
\usepackage{todonotes}
\usepackage{menukeys}
\usepackage{hyperref}
\usepackage{acronym}
\usepackage{array,mathtools,amssymb,booktabs}
\usepackage{physics}
\AtBeginDocument{
\heavyrulewidth=.10em
\lightrulewidth=.05em
\cmidrulewidth=.03em
\belowrulesep=.65ex
\belowbottomsep=0pt
\aboverulesep=.4ex
\abovetopsep=0pt
\cmidrulesep=\doublerulesep%
\cmidrulekern=0.5em
\defaultaddspace=.5em
}

\definecolor{mygray}{RGB}{128,128,128}

\usepackage{acronym}

\begin{document}
\newlength\figurewide%
\ifFIGoneColumn%
  \figurewide=.5\columnwidth%
\else
  \figurewide=.9\columnwidth%
\fi

\title{Phase-space resolved rates in driven multidimensional chemical reactions}
\author{Matthias Feldmaier}
\author{Robin Bardakcioglu}
\author{Johannes Reiff}
\author{J\"org Main}
\affiliation{%
Institut f\"ur Theoretische Physik 1,
Universit\"at Stuttgart,
70550 Stuttgart,
Germany}

\author{Rigoberto Hernandez}
\email[Correspondence to: ]{r.hernandez@jhu.edu}
\affiliation{%
Department of Chemistry,
Johns Hopkins University,
Baltimore, Maryland 21218, USA}
\date{\today}
\renewcommand{\vb}[1]{\boldsymbol{#1}}
\renewcommand{\vec}[1]{\boldsymbol{#1}}
\newcommand{\Ws}{\mathcal{W}_\mathrm{s}}
\newcommand{\Wu}{\mathcal{W}_\mathrm{u}}
\newcommand{\Wsu}{\mathcal{W}_\mathrm{s,u}}
\newcommand{\sno}[1]{_\mathrm{#1}}
\newcommand{\EQ}{Eq.\ }
\newcommand{\EQS}{Eqs.}
\newcommand{\FIG}{Fig.}
\newcommand{\FIGS}{Figs.}
\newcommand{\REF}{Ref.}
\newcommand{\REFS}{Refs.}
\newcommand{\SEC}{Sec.}
\newcommand{\SECS}{Secs.}
\newcommand{\eg}{e.\,g.}
\newcommand{\ie}{i.\,e.}
\newcommand{\cf}{cf.}
\newcommand{\TODO}[1]{{\color{red}TODO: #1}}
\newcommand{\no}[1]{\mathrm{#1}}

\begin{abstract}
Chemical reactions in multidimensional driven systems are typically
described by a time-dependent rank-1 saddle associated with one reaction and
several orthogonal coordinates (including the solvent bath).
To investigate reactions in such systems,
we develop a fast and robust method
---viz., local manifold analysis (LMA)---
for computing the instantaneous decay rate of reactants.
Specifically, it computes the instantaneous decay rates along
saddle-bound trajectories near the activated complex
by exploiting local properties of the stable and unstable manifold
associated with the normally hyperbolic invariant manifold (NHIM).
The LMA method
offers substantial reduction of numerical effort and increased
reliability in comparison to
direct ensemble integration.
It provides an instantaneous
flux that can be assigned to every point on the NHIM
and which is associated with a
trajectory---regardless of whether it is periodic, quasi-periodic, or
chaotic---that is bound on the NHIM.
The time average of these fluxes in the driven system corresponds
to the average rate through a given local section
containing the corresponding point on the NHIM.
We find good agreement between the results of the LMA
and direct ensemble integration
obtained using numerically constructed, recrossing-free dividing surfaces.
\end{abstract}
\maketitle

\preto\section\acresetall
\acrodef{NN}{neural network}
\acrodef{DS}{dividing surface}
\acrodef{TST}{Transition State Theory}
\acrodef{TS}{Transition State}
\acrodef{LD}{Lagrangian descriptor}
\acrodef{LMA}{local manifold analysis}
\acrodef{NHIM}{normally hyperbolic invariant manifold}
\acrodef{PSOS}{Poincar\'e surface of section}

\section{Introduction}
\label{sec:Introduction}
The dynamics of a chemical system can often be described by
the classical equations of motion for selected coordinates driven
by a Born-Oppenheimer potential.
The transition from reactants to products in a such a chemical
reaction is typically marked by a barrier region with a rank-1 saddle
that has exactly one unstable direction called the reaction
coordinate, while the remaining degrees of freedom are locally stable
and are associated with other bound internal motions and
external bath coordinates.
The reaction can be described both qualitatively and quantitatively
within the framework of \ac{TST}. 
This theory rests on the identification of a recrossing-free \ac{DS} in
the barrier region, which separates reactants from
products,\cite{eyring35,wigner37,pech81,truh96,peters14a,wiggins16}
and has been applied in a  wide range of fields,
as we have noted in our previous work (see e.g., in Ref.~\citenum{hern19a}),
and in recent advances in chemical reaction
dynamics\cite{lorquet17,keshavamurthy18,waalkens18,komatsuzaki18}.

The construction of recrossing-free \acp{DS} is a non-trivial task,
that can be achieved by using perturbation theory to construct
good action-angle variables associated with the \ac{DS}.\cite{hern93b,wiggins01,Uzer02}
The theory was generalized for chemical reactions under
time-dependent conditions, arising from driving, noise, or
both,\cite{dawn05a,hern08d} and has been applied to chemical reactions
with few degrees of freedom including, e.g., H $+$
H$_2$,\cite{hern94,Allahem12} LiCN,\cite{hern16c} and ketene
isomerization.\cite{hern16d}

Typical chemical reactions, however, require representations with
higher dimensionality coupled to complex environments.
In a multidimensional autonomous Hamiltonian system, a recrossing-free
\ac{DS} attached to the \ac{NHIM}
contains all trajectories that are trapped in the saddle region when
propagated both forward and backward in time.\cite{hern17h,hern19a}
This manifold can be constructed approximately using normal form
expansions,\cite{pollak78,pech79a,hern93b,hern94,Uzer02,Jaffe02,
Teramoto11,komatsuzaki06a,Waalkens04b,Waalkens13}, or exactly
by numerical application of Lagrangian descriptors,\cite{hern17h,hern19a}
the binary contraction method,\cite{hern18g} and machine learning
algorithms.\cite{hern18c,hern19a}

In a driven system with one degree of freedom,
the rate constant can
be obtained by propagating a large ensemble of trajectories in the
vicinity of the \ac{TS} trajectory ---which coincides with the
time-dependent \ac{NHIM} in a one-dimensional system---
and observing the decay in
the number of reactant trajectories, which have
not yet crossed the \ac{DS}.
As a numerically less expensive alternative, the rate constant can be
computed using the Floquet exponents of the \ac{TS} trajectory.\cite{hern14f}
However, the situation is less clear for systems with $d \ge 2$
degrees of freedom,
because the dimensionality of the time-dependent NHIM
is $(2d-2) > 0$ and hence there is not a unique trajectory from
which to obtain a single set of stability exponents.
Indeed, the additional degrees of freedom enable a non-trivial
dynamics on the \ac{NHIM},\cite{hern19T1} and substantially extends the
possibilities for trajectories to cross a \ac{DS} close to the \ac{NHIM}.

The purpose of this paper is to address this challenge in
obtaining rate constants in
multidimensional driven systems with a rank-1 saddle
coupled to a formally arbitrary number of orthogonal modes.
We find that we can assign instantaneous
reaction rates to every point on the time-dependent
\ac{NHIM}
by observing the change in
the number of trajectories crossing the numerically constructed
time-dependent \acp{DS}.
We also address the question of how appropriate initial
conditions of trajectory ensembles must be chosen to compute the instantaneous
rate through a local section of the \ac{NHIM}.
    Trajectories having just enough energy
    to overcome the time dependent barrier
    will cross the \ac{DS} close to the \ac{NHIM}.
    They correspond to the slowest flux and are therefore critical
    in determining instantaneous rates,
    which are approximately given by the decay rates of the \ac{TS},
    at least in the local vicinity of the \ac{NHIM}.
    The time averages of these instantaneous rates
yield rate constants associated with periodic,
quasi-periodic, or even chaotic trajectories on the \ac{NHIM}.
Furthermore, we present a very efficient numerical method for the
computation of instantaneous reaction rates associated to arbitrary
points on the \ac{NHIM}, which does not require the propagation of
trajectory ensembles.
This \ac{LMA} method uses local properties of the stable and unstable
manifolds at points on the \ac{NHIM}.
Reaction rates calculated using the \ac{LMA} method
agree with those
obtained through the more numerically
expensive sampling of the trajectory ensemble.

The paper is organized as follows.
In Sec.~\ref{sec:Theory}, we introduce our methods for the computation
of reaction rates based
on the propagation of trajectory ensembles
in Sec.~\ref{sec:ensemble_rates},
on the Floquet exponents of trajectories on the \ac{NHIM}
in Sec.~\ref{sec:floquet_method},
and
on local properties of the stable and unstable manifolds at points on the \ac{NHIM}
in Sec.~\ref{sec:manifold_method}.
Numerical results and comparisons are presented in Sec.~\ref{sec:Results}.
As this work is focused on advancing methods so as to ultimately
calculate rate constants of multidimensional chemical reactions,
numerical results are restricted to a periodically driven
two-dimensional model system useful for verifying the theory for
periodic and general non-periodic trajectories in
Secs.~\ref{sec:periodic_trajectories} and \ref{sec:non-periodic_trajectories},
respectively.

\section{Theory and methods}
\label{sec:Theory}
\begin{figure}
  \includegraphics[width=0.9\columnwidth]{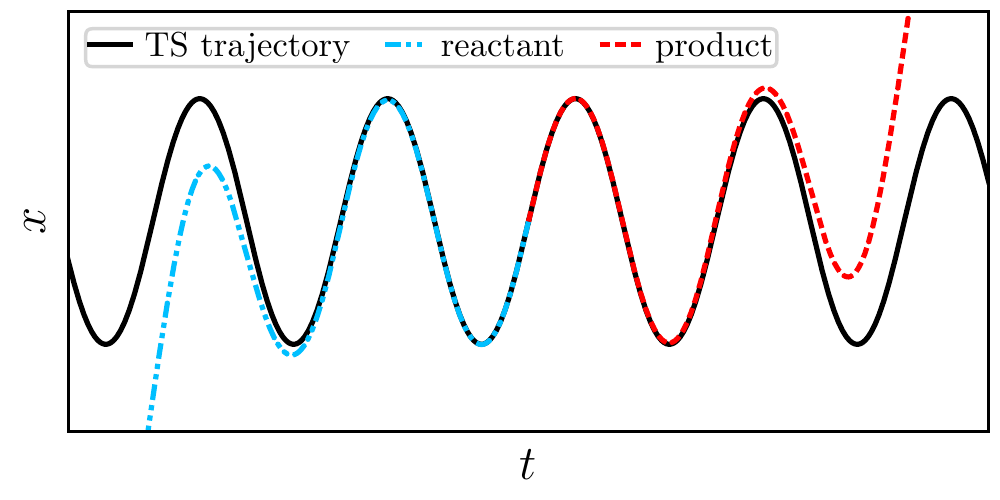}
  \caption{Sketch of a reacting trajectory moving close to the \ac{TS}
    trajectory (black solid line) in the barrier region.  The
    trajectory changes from reactant (blue dash-dotted line) to
    product (dashed red line) when crossing the \ac{DS}.}
  \label{fig:TS_sketch}
\end{figure}

In this section we investigate rate constants for a driven chemical
reaction in a system with $d \ge 2$ degrees of freedom.
More precisely, the system is described by a moving rank-1 saddle with
one unstable reaction coordinate $x$ and $d-1$ stable bath coordinates
$\vec y$.
The calculation of rate constants in \ac{TST} usually implies that the
pathway of reactive trajectories is located close to the \ac{TS}, as
illustrated in Fig.~\ref{fig:TS_sketch}.
Exceptions are roaming reactions\cite{bowman04a,hern13e,wiggins14a,
bowman2011c,bowman14a} with pathways far away from the \ac{TS}, which
we do not consider in this paper.
We also do not investigate the Kramers rate\cite{rmp90} for an
ensemble of trajectories prepared in a reactant well far away from
the saddle.
While the Kramers rate depends on the thermally distributed
kinetic energy of the trajectories in the well, in this paper the term
\emph{rate} refers to the decay rate of
individual trajectories in the activated complex near the \ac{TS},
which is directly related to geometrical and dynamical
properties of the barrier itself, but not
necessarily to properties of the reacting ensemble directly.

The effect of the barrier and its driving on reaction rates is high
for trajectories that react very closely to the \ac{TS} and therefore
stay in the saddle region for a long time.
In one-dimensional static systems with an energy barrier, the
point-like \ac{TS}
defines the minimum energy required for a reaction.
It also yields the precise separation between reactants and products,
and therefore a recrossing-free \ac{DS} can be attached to this point.
If the system is subject to time-dependent driving, this point becomes
the one-dimensional \ac{TS} trajectory, and a time-dependent
recrossing-free \ac{DS} can be attached to the \ac{TS} trajectory to
separate reactants from products.\cite{hern15a}

In multidimensional systems with a time-dependent rank-1 saddle the
situation is more complex.
The \ac{NHIM}
becomes a higher-dimensional
manifold and is no longer a single \ac{TS} trajectory.
Consequently, reacting trajectories have additional degrees of
freedom associated with the crossing
of the \ac{DS} close to the \ac{NHIM}.
In Ref.~\citenum{hern19T1}, the reaction rate for a specific periodic
trajectory of a higher-dimensional system has been obtained using the
Floquet exponents of that trajectory.
Here, we present three methods to calculate reaction rates
either using or inspired by this trajectory-based approach
focused on the flux through arbitrary points on the \ac{NHIM}.
The first method discussed in Sec.~\ref{sec:ensemble_rates} is based
on the propagation of trajectory ensembles with appropriately chosen
initial conditions.
The second, numerically less expensive method presented in
Sec.~\ref{sec:floquet_method} is an extension of the Floquet method of
Ref.~\citenum{hern14f} to multidimensional systems.
The third method derived in Sec.~\ref{sec:manifold_method} uses local
properties of the stable and unstable manifolds for the computation of
the instantaneous reaction rates.

\subsection{Ensemble method}
\label{sec:ensemble_rates}
\subsubsection{Preparation of trajectory ensembles}
\label{sec:ensemble_preparation}
Reaction rates are usually obtained by propagating a large number of
trajectories
and by measuring their flux from a reactant to a product state.
When propagating ensembles, the
accurate preparation in the reactant state
is crucial to obtaining conclusive results for the instantaneous rate.
As these rates are a property of the (possibly time-dependent) saddle,
it is important that an ensemble is prepared in a way that many trajectories
are affected by the saddle.
If most of the trajectories would react at an energy
much higher than the barrier, they will most likely cross the barrier region
very fast
and their dynamics will be mostly unaffected by the saddle.
Because of this, it is important to initialize an ensemble close to
the stable manifold, where trajectories closely approach the \ac{NHIM} and are
therefore strongly affected by the saddle-bound dynamics on the \ac{NHIM}.

\begin{figure}
  \includegraphics[width=0.95\columnwidth]{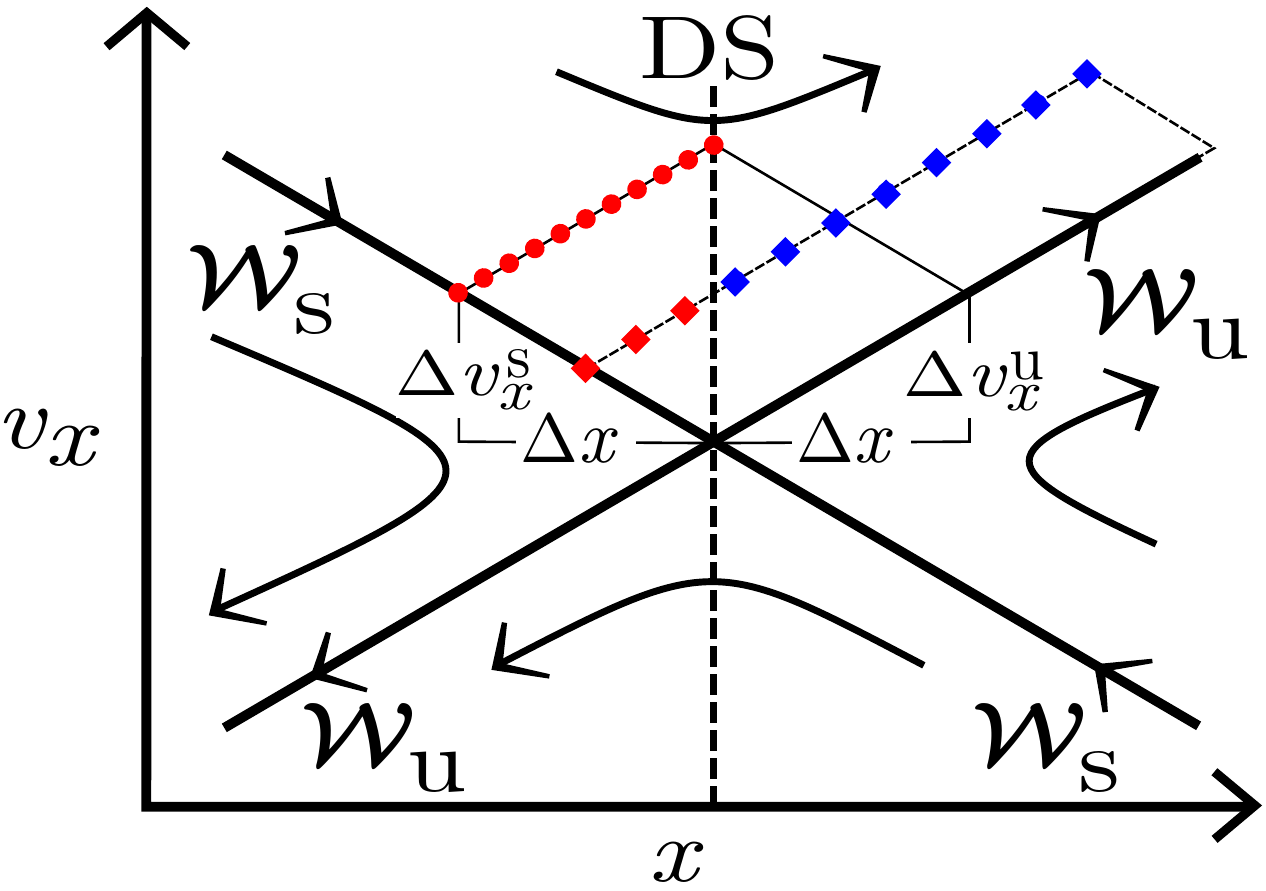}
  \caption{Sketch of the stable and unstable manifolds $\mathcal{W}\sno{s,u}$
    in the $(x,v_x)$ plane at fixed parameters $(\vec{y},\vec{v}_y, t)$
    of bath coordinates and time of a multidimensional driven
    system. The intersection of the two manifolds is a point
    $(x,v_x)^{(\mathrm{NHIM})}(\vec{y},\vec{v}_y, t)$ on the
    time-dependent \ac{NHIM}. The \ac{DS} attached to
    this point is marked by a vertical dashed line. The arrows depict the
    four reactive and non-reactive regions separated by the manifolds.
    The red points mark the initial conditions of a reactant trajectory ensemble
    at time $t$ arranged equidistantly along a line segment parallel
    to the unstable manifold. The diamonds indicate the ensemble at a
    later time $t+\Delta t$, where this ensemble has partially crossed
    the \ac{DS} (blue diamonds).
}
  \label{fig:ensemble_preparation}
\end{figure}
In Fig.~\ref{fig:ensemble_preparation} a schematic of the stable $\mathcal{W}
\sno{s}$ and unstable manifold $\mathcal{W}\sno{u}$ in a $(x, v_x)$ section of
a multidimensional system is shown. The limits of this section are small enough,
such that the dynamics close to the \ac{NHIM} are linear,
and this requires $\Delta x$ to also be small.
We attach a recrossing-free \ac{DS} to the \ac{NHIM}, which is shown as a
vertical, dashed line through the intersection of $\mathcal{W}\sno{s}$
and $\mathcal{W}\sno{u}$ in Fig.~\ref{fig:ensemble_preparation}.

As highlighted in Fig.~\ref{fig:ensemble_preparation} by the red
dots,
an ensemble of $N_{\sno{react}}(0)$ equidistant reactants is initialized close
to the \ac{NHIM}, on a line segment parallel to $\mathcal{W}\sno{u}$,
spanning from a point on $\mathcal{W}\sno{s}$ to the \ac{DS}.
If a system is time-dependent, the ensemble preparation
procedure can be repeated for any point on the NHIM at a time $t$.
How often such an ensemble needs to be prepared and at which initial times $t$
is determined by the rate calculation,
is explained in Sec.~\ref{sec:ensemble_rate}.

Further discussion on the \ac{NHIM} and recrossing-free \acp{DS} are found in
Ref.~\citenum{hern19a} and references therein. Algorithms to find the \ac{NHIM}
or to find $\mathcal{W}\sno{s}$, as well as the construction of the \ac{DS} are
 presented in Refs.\citenum{hern19a,hern18g,hern17h}.

\subsubsection{Instantaneous ensemble rates}
\label{sec:ensemble_rate}
Having prepared an ensemble of reactive trajectories
according to Sec.~\ref{sec:ensemble_preparation},
these trajectories need to be numerically propagated in time. Starting with a
reactant
population of $N\sno{react}(0)$,
each time a trajectory reacts or, more precisely, pierces the recrossing-free
DS, the reactant population
decreases by one. For a system that is not time-dependent,
the number of reactants $N\sno{react}$ will decrease exponentially fast
\begin{equation}
	N\sno{react}(t) \propto \exp(- k\,t)~,
	\label{fit_rate_constant}
\end{equation}
with $k$ being the rate constant of this reaction.
If a barrier is time-dependently driven,
depending on the driving itself as well as on the initial time $t$
an ensemble is prepared (here, e.\,g.\ for periodically driven systems,
the phase relative to the barrier's oscillation is important),
the decrease of $N\sno{react}(t)$ will be more or less exponential.
So fitting a rate constant $k$ according to Eq.~\eqref{fit_rate_constant}
may be approximately possible at best in most cases.
If chosen wrong, the parameter $\Delta x$ introduced in
Sec.~\ref{sec:ensemble_preparation} when preparing an ensemble will also have a huge
influence on the decay of reactants.

The problem of obtaining a precise rate for
a non-exponentially decaying reactant population $N\sno{reac}(t)$
can be solved by using a more fundamental definition of rates.
In the framework of chemical kinetics,%
\cite{upadhyay06,connors90,Haenggi1990}
the change in a reactant's population
in first order, unimolecular reactions is proportional
to the reactant's population
\begin{equation}
  \frac{\no{d}}{\no{d}t} N\sno{react}(t) = -k(t) N\sno{react}(t)~,
  \label{rate_equation_1}
\end{equation}
where we allow $k(t)$ to be time-dependent.
In the time-independent case, Eq.~\eqref{rate_equation_1}
would be solved by Eq.~\eqref{fit_rate_constant}.
However, the instantaneous rate $k(t)$ yields
\begin{equation}
    k(t) = -\frac{1}{N\sno{react}(t)} \frac{\no{d}}{\no{d}t} N\sno{react}(t)
    = - \frac{\no{d}}{\no{d}t}\ln(N\sno{react}(t))~.
    \label{time_dependent_rate}
\end{equation}

\begin{figure}
  \includegraphics[width=0.9\columnwidth]{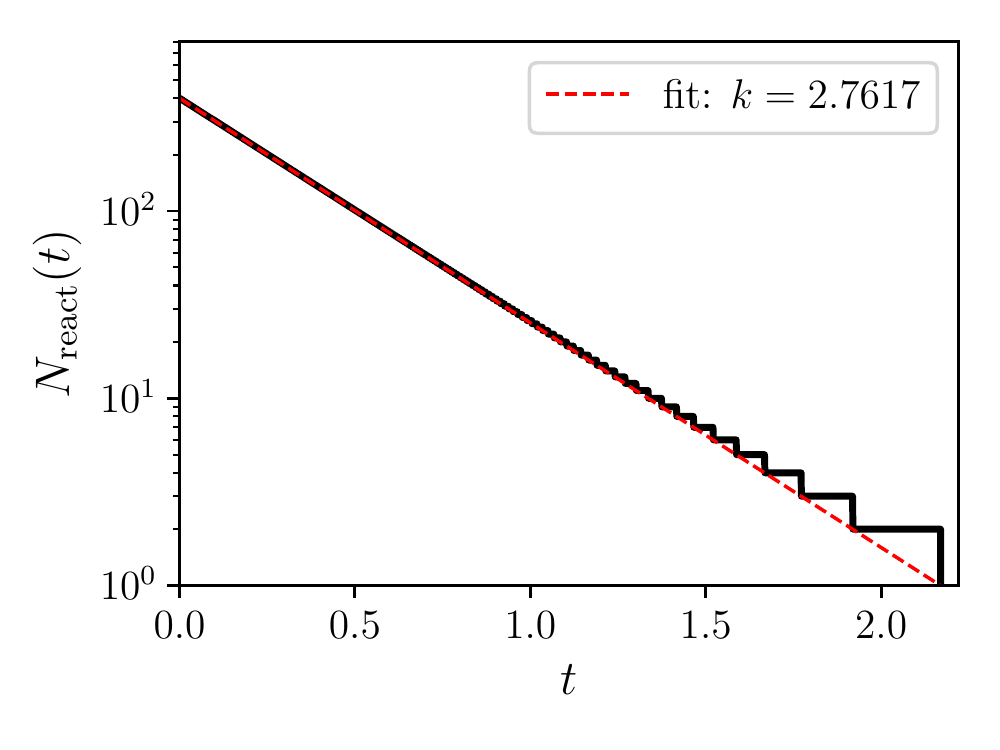}
  \caption{Rate constant for the TS of the static two-dimensional model system
    according to Eq.~\eqref{eq:potential} with $\hat x = 0$.
	An ensemble of $400$ reactive trajectories
	was prepared with $\Delta x = 10^{-3}$ according to Fig.~\ref{fig:ensemble_preparation}
	close to the TS and dynamically propagated. Each time one of the trajectories
	reacts, the number of reactants $N\sno{react}(t)$ is decreased by one. The
	red dashed line shows a fit according to Eq.~\eqref{fit_rate_constant} to the
	reaction events, yielding a rate constant of $k=2.7617$. The step-function
	shown as a black solid line as a guide to the eye is to visualize,
	that not only $N\sno{react}(t)$ but also the density of reaction events decreases
	exponentially fast.}
  \label{fig:hist_static_rate}
\end{figure}

Due to the way we measure $N\sno{react}(t)$, an exponential decay of reactants
also translates into an exponential decay of measurement density
for $N\sno{react}(t)$
over time, as seen in the step function behavior
in Fig.~\ref{fig:hist_static_rate}.
If the time intervals between data points are too small or too large,
numerical methods of differentiation
may produce erroneous results, which has to be considered during evaluation.

Since the decay is exponential, the number of trajectories necessary to reduce
the time intervals between individual reaction events, i.e.\ to increase the
time range where the differentiation of $N\sno{react}(t)$ produces accurate
results, grows exponentially. Hence, increasing the number of trajectories for
a single ensemble will produce diminishing results.
For this reason, we start several ensembles
at different initial times $t$ and piece together the results
of each ensemble to fully resolve $k(t)$ of a specific trajectory on the
\ac{NHIM} with good accuracy everywhere.

Whereas in other references, thousands to millions of trajectories had to be
propagated
to somehow resolve a rate constant,\cite{hern19a,hern18c,hern17h,hern17f}
with the methods introduced here we are able to adequately resolve
the time-dependent reaction rate $k(t)$ propagating just a few hundreds of
trajectories.

\subsection{Floquet method}
\label{sec:floquet_method}
Propagating an ensemble of trajectories to obtain reaction rates is
numerically expensive as these ensembles, usually with a large number
of reactive trajectories, have to be propagated in time.
An alternative method to obtain reaction rates, which does not require
the propagation of a trajectory ensemble has been introduced in 2014
by Craven et~al.\ for one-dimensional systems with periodic driving.\cite{hern14f}
The method is based on the stability analysis of the \ac{TS}
trajectory yielding the \emph{Floquet rate}
\begin{equation}
  \overline{k}\sno{F} = \mu\sno{u} - \mu\sno{s}
  \label{eq:floquet_rate}
\end{equation}
which depends on the Floquet exponents
\begin{equation}
  \mu\sno{u,s} = \frac{1}{T}\,\ln |m\sno{u,s}(T)|
  \label{eq:floquet_exponents_periodic}
\end{equation}
of that trajectory. Here, $m\sno{u,s}$ are the eigenvalues of the
$(2\times 2)$-dimensional stability (or monodromy) matrix along the
unstable and stable directions, respectively, and $T$ is the period of the external driving.
The method has already been extended and applied by Tsch\"ope et~al.\
to the \ac{TS} trajectory of a higher-dimensional system with
periodic driving.\cite{hern19T1}.
In a system with $d$ degrees of freedom $m\sno{u,s}$ in
Eq.~\eqref{eq:floquet_exponents_periodic} are the two eigenvalues of
the $(2d\times 2d)$-dimensional stability matrix of the \ac{TS}
trajectory, or more general, of any periodic orbit with any period $T$,
with the highest and lowest absolute values.

For a first order, two-dimensional differential equation
$\dot{\vec{\gamma}} = \vec{f}(\vec{\gamma}, t)$, where
$\vec{\gamma} \in R^{2d}$ is a vector in phase space,
the stability matrix $\vec{M}(t)$ is obtained through linearization of the
dynamics, which is in turn characterized by the Jacobian
$\vec{J}(\vec{\gamma}(t),t) =  \partial\vec{f} / \partial\vec{\gamma}$ of the system
\begin{equation}
    \dot{\vec M} = \vec J \vec M\, ,
    \label{eq:mdot}
\end{equation}
with initial conditions $\vec{M}(0) = \vec{1}_{2d}$.

For a Hamiltonian system with time-dependent potential $V(\vec{q},t)$
and without friction, Eq.~\eqref{eq:mdot} takes the form
\begin{equation}
  \dot{\vec M} =
  \begin{pmatrix}\vec{0}_d & \vec{1}_d\\
    -\frac{\partial^2 V}{\partial \vec{q}^2} & \vec{0}_d\end{pmatrix} \vec{M} \, .
  \label{eq:mdot-hamiltonian}
\end{equation}
In Hamiltonian systems, the monodromy matrix $\vec{M}$ is symplectic, and
hence, if $\lambda$ is an eigenvalue of $\vec{M}$, then also $1/\lambda$
is an eigenvalue, as well as their complex conjugates $\bar\lambda$ and
$1/\bar\lambda$.

The Floquet method can also be used to calculate the reaction rates
of non-periodic trajectories on
the \ac{NHIM}.
In that case Eq.~\eqref{eq:floquet_exponents_periodic} must be
replaced with
\begin{equation}
  \mu\sno{u,s} = \lim_{t\to\infty} \frac{1}{t}\,\ln|m\sno{u,s}(t)| \, .
  \label{eq:floquet_rate_non_periodic}
\end{equation}
In numerical simulations the time $t$ must be chosen sufficiently large
to obtain converged estimates of the Floquet rate $\overline{k}\sno{F}$ in
Eq.~\eqref{eq:floquet_rate}.

The Floquet method described in Ref.~\citenum{hern14f} can only be
used to obtain mean rate constants averaged over one period of a
periodic trajectory or obtained in the limit $t\to\infty$, but can not
provide the instantaneous rates $k(t)$ discussed in Sec.~\ref{sec:ensemble_rate}.
The reason is that the directions of the stable and unstable manifolds
associated to points on the \ac{NHIM} change with time, and thus, in
general can not be obtained as eigenvectors of the stability matrix.

\subsection{Local manifold analysis}
\label{sec:manifold_method}
Consider an initial ensemble of trajectories in the local vicinity of
the stable and unstable manifolds of an arbitrary point on the
\ac{NHIM} in the same way as described in Sec.~\ref{sec:ensemble_preparation},
 sketched by the red circles in Fig.~\ref{fig:ensemble_preparation}.
The idea is not to propagate the individual trajectories of the
ensemble by numerical integration of the equations of motion, but to
use the linearized dynamics in the local vicinity of the stable and
unstable manifold to propagate the whole line segment.
When going from time $t$ to time $t + \Delta t$ the linearized dynamics
can be split into two parts. A compression towards the \ac{NHIM} in the
direction of the stable manifold by the factor
$\Delta x^{\no{s}}(t+\Delta t)/ \Delta x^{\no{s}}(t)$
pulls the line closer to the unstable manifold and a stretching in the
direction of the unstable manifold by the factor
$\Delta x^{\no{u}}(t+\Delta t)/ \Delta x^{\no{u}}(t)$ lengthens the line
segment. The result of this motion is shown by the red and blue diamonds in
Fig.~\ref{fig:ensemble_preparation}. With simple geometry, we can now determine
the ratio between the length of the line segment left of the DS (red diamonds
in Fig.~\ref{fig:ensemble_preparation}) and the length of the entire line. Due
to the properties of the linear mapping that occurred, we know that the
density of reactive trajectories along the line is constant.
Thus, the obtained ratio of line
segments is proportional to the number of reactants $N\sno{react}(t)$.
By choosing $\Delta x^{\no{s}}(t)=\Delta x^{\no{u}}(t)\equiv\Delta x$
as marked in Fig.~\ref{fig:ensemble_preparation}, the number of
reactants at time $t + \Delta t$ is given as
\begin{equation}
  N\sno{react}(t+\Delta t) =
  \frac{\Delta x^{\no{s}}(t+\Delta t)}{\Delta x^{\no{u}}(t+\Delta t)} N_{\no{react}}(t) \; .
\label{eq:N_t}
\end{equation}
It is important to note that the stable and unstable manifolds shown
in Fig.~\ref{fig:ensemble_preparation} can independently rotate around their
intersection point as functions of time. However, Eq.~\eqref{eq:N_t} stays
valid
in these cases. One can always separate the linear mapping that occurs into two
mappings. First, one that transforms the parallelogram depicted in Fig.~\ref
{fig:ensemble_preparation}, such that these rotations are accounted for, all
the while keeping $\Delta x^{\no{u}}(t)$ and $\Delta x^{\no{s}}(t)$ constant,
which in turn ensures that the line segment does not cross further
into the DS before the next step.
Subsequently, we can
perform the stretching and compression of the appropriate lengths, as
discussed before, to
complete the mapping of the linearized dynamics.

From Eq.~\eqref{eq:N_t} we obtain the instantaneous rate
\begin{align}
  k(t) &= - \frac{\no d}{\no d t}\ln N\sno{react}(t) =
  \frac{\no d}{\no d t}
  \left[\ln\Delta x^{\no{u}}(t) - \ln\Delta x^{\no{s}}(t) \right] \nonumber \\
       &= \frac{\Delta\dot x^{\no{u}}(t)}{\Delta x^{\no{u}}(t)}
        - \frac{\Delta\dot x^{\no{s}}(t)}{\Delta x^{\no{s}}(t)} \, .
\label{eq:k_t_prel}
\end{align}
Using the linearized equations of motion \eqref{eq:mdot-hamiltonian} for the
stability matrix $\vec M$, it can be shown that
$\Delta\dot x^{\no{s}}(t)=\Delta v_x^{\no{s}}(t)$ and
$\Delta\dot x^{\no{u}}(t)=\Delta v_x^{\no{u}}(t)$.
Inserting this into Eq.~\eqref{eq:k_t_prel} and also explicitly
writing the dependencies on the bath coordinates $\vec y$ and
velocities $\vec{v}_y$, which have been dropped in the above
derivations, we finally obtain
\begin{equation}
  k(\vec{y},\vec{v}_y,t)
  = \frac{\Delta v_x^{\no{u}}}{\Delta x^{\no{u}}}(\vec{y},\vec{v}_y,t)
  - \frac{\Delta v_x^{\no{s}}}{\Delta x^{\no{s}}}(\vec{y},\vec{v}_y,t) \, .
\label{eq:k_t_final}
\end{equation}
This means that for any point on the \ac{NHIM} parameterized by the
bath coordinates $\vec y$ and velocities $\vec{v}_y$ the
instantaneous rate $k(\vec{y},\vec{v}_y,t)$ is simply given by the
difference of the slopes of the stable and unstable manifolds in
the $(x,v_x)$ diagram shown in Fig.~\ref{fig:ensemble_preparation}. Note that
for non-Hamiltonian systems, $\Delta \dot x^{\no{u,s}}$ depend on the
contents of the Jacobian $\vec J$ in
Eq.~\eqref{eq:mdot}, which, in general,
emerges as a prefactor to the slopes in Eq.~\eqref{eq:k_t_final}. To
determine the slope, one can determine the positions of these manifolds in the
immediate vicinity of the \ac{NHIM} with
methods discussed in Ref.~\citenum{hern18g}, where stable and unstable manifolds are
interpreted as boundaries between reactive and non-reactive regimes.

In the derivation of the \ac{LMA} method, we have assumed that the
position of the \ac{DS} is the same for all particles of the
propagated ensemble at time $t+\Delta t$ (see the diamonds in
Fig.~\ref{fig:ensemble_preparation}).
This assumption is satisfied in the system discussed in
Sec.~\ref{sec:Results} but may not be so in arbitrary systems.
For example, in some multidimensional systems the orthogonal modes
may be coupled to the velocity $v_x$ of the reaction coordinate, and
thus propagated positions will be associated with different values
$(\vec{y},\vec{v}_y)$ of the DS
$x_{\mathrm{DS}}(\vec{y},\vec{v}_y,t+\Delta t)$.
Addressing such challenging cases is a subject for future work.

\section{Results and discussion}
\label{sec:Results}
To verify the methods described in Sec.~\ref{sec:Theory} we use the
same model for a driven chemical reaction
investigated earlier.\cite{hern17h,hern18c,hern18g,hern19a,hern19T1}
The system is described by the two-dimensional potential
\begin{align}
  V(x, y, t) &= E\sno{b}\,\exp\left(
  -{\left[x- \hat{x} \sin\left(\omega_x t \right)\right]}^2\right) \nonumber \\
&+ \frac{\omega_y^2}{2}{\left[y-\frac{2}{\pi} \arctan\left(2 x \right)\right]}^2 \; ,
\label{eq:potential}
\end{align}
where a time-dependent oscillating Gaussian barrier with height
$E\sno{b}$ separates an open reactant from an open product basin.
The barrier is moving along the $x$-coordinate with frequency
$\omega_x$ and amplitude $\hat x$.
In the $y$-direction, the dynamics is bound by a harmonic potential
with frequency $\omega_y$.
The $x$ and the $y$ coordinates are non-linearly coupled, so that the
minimum energy path for a reaction over the saddle is given by
$y=(2/\pi)\,\no{arctan}(2x)$.
For simplicity, we use dimensionless units in which the parameters are
$E\sno{b}=2$, $\omega_x =\pi$, $\omega_y=2$ and $\hat x = 0.4$ if not
stated otherwise.
The open reactant and product basins of the potential \eqref{eq:potential}
allow us to study different methods on how to obtain reaction rates
without having to worry about global recrossings, that arise
if one or both basins are closed.\cite{hern17e}
However, open reactant and product basins are not a prerequisite for
the application of the methods presented here, as these methods are
based on the local dynamics near the \ac{NHIM}.

\begin{figure}
  \includegraphics[width=\columnwidth]{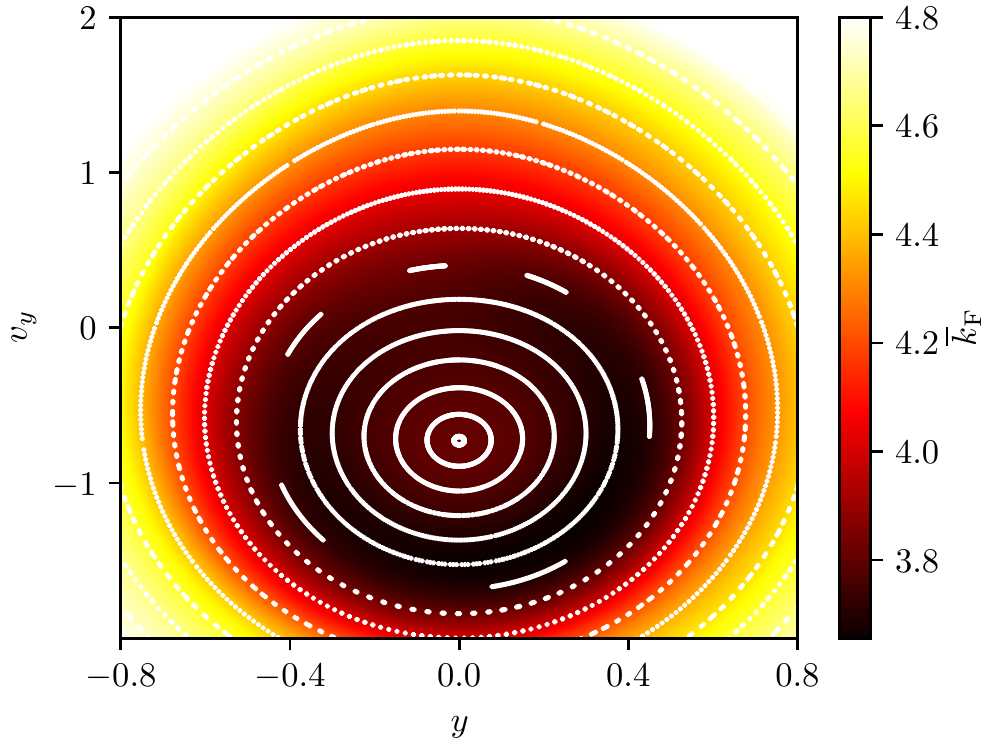}
	\caption{Floquet rates $\overline{k}\sno{F}$ obtained for trajectories initialized at
  $t=0$ on the corresponding NHIM of the time-periodically driven potential~\eqref{eq:potential}.
  The rate is visualized as color encoding on the $(y, v_y)$-surface
  representing the coordinates on the NHIM.
  The white dots represent a PSOS, where selected trajectories are propagated
  for many oscillation periods of the potential. Their
  instantaneous positions $y(t=n\,T)$ and velocities $v_y(t=n\,T)$
  with $n\in \mathbb{N}$ at integer
  multiples of the potential's oscillation period $T=2$ are marked with these white dots,
  indicating the stable tori of the dynamics around the \ac{TS} trajectory, that is located
  in the center of the tori.
}
  \label{fig:floquet_nhim}
\end{figure}

The \ac{NHIM} of the model system \eqref{eq:potential} is a
two-dimensional time-dependent manifold.
The dynamics of trajectories on the \ac{NHIM} has been studied in
Ref.~\citenum{hern19T1}.
For a periodically driven system with $d=2$ degrees of freedom, it
can be visualized by a \ac{PSOS}, e.g., the points $(y,v_y)(t=nT)$
obtained by using a stroboscopic map in time, with $T=2$ the period of
the driving and $n\in\mathbb{N}$.
The \ac{PSOS} for the system \eqref{eq:potential} is presented by the
white points in Fig.~\ref{fig:floquet_nhim} and exhibits regular
torus-like structures.
The fixed point with coordinates $y=0$, $v_y=-0.72$ located at the
center of the tori corresponds to a periodic trajectory with the same
period as the external driving, and we call this orbit the \ac{TS}
trajectory.
Trajectories on the surrounding tori typically show a quasi-periodic
behavior.
As discussed in Sec.~\ref{sec:Theory}, rate constants can be assigned
to trajectories on the \ac{NHIM}.
This is illustrated for periodic and arbitrary non-periodic trajectories
on the \ac{NHIM} in the following Secs.~\ref{sec:periodic_trajectories}
and \ref{sec:non-periodic_trajectories}, respectively.

\subsection{Rate calculation for periodic trajectories on the NHIM}
\label{sec:periodic_trajectories}
\begin{figure}
  \includegraphics[width=0.95\columnwidth]{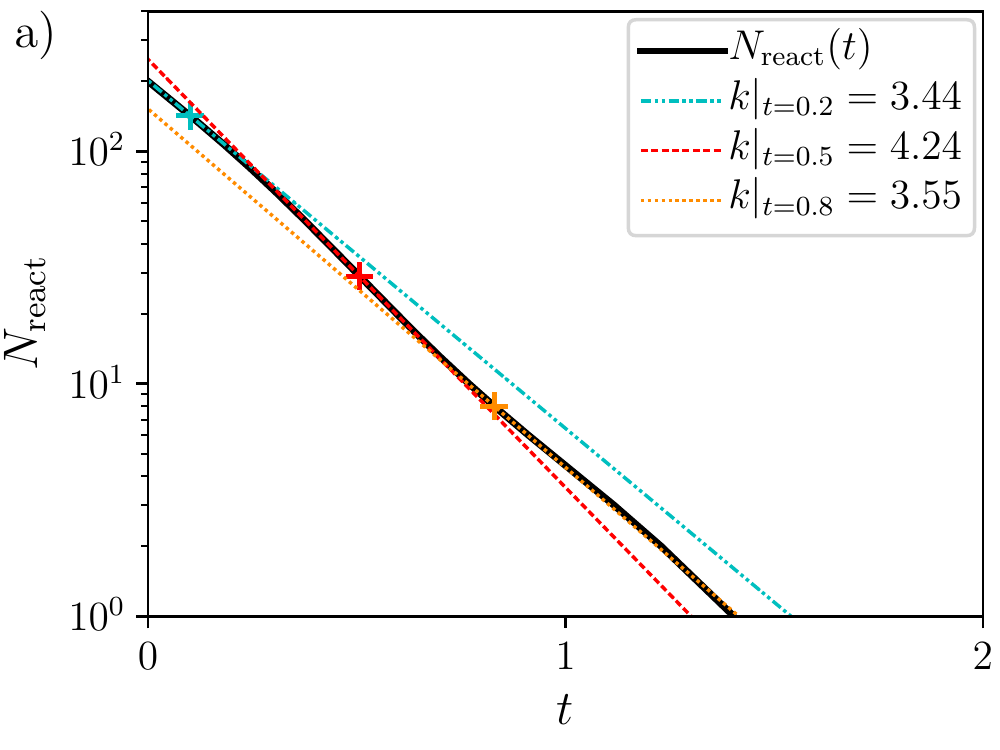}
  \includegraphics[width=0.95\columnwidth]{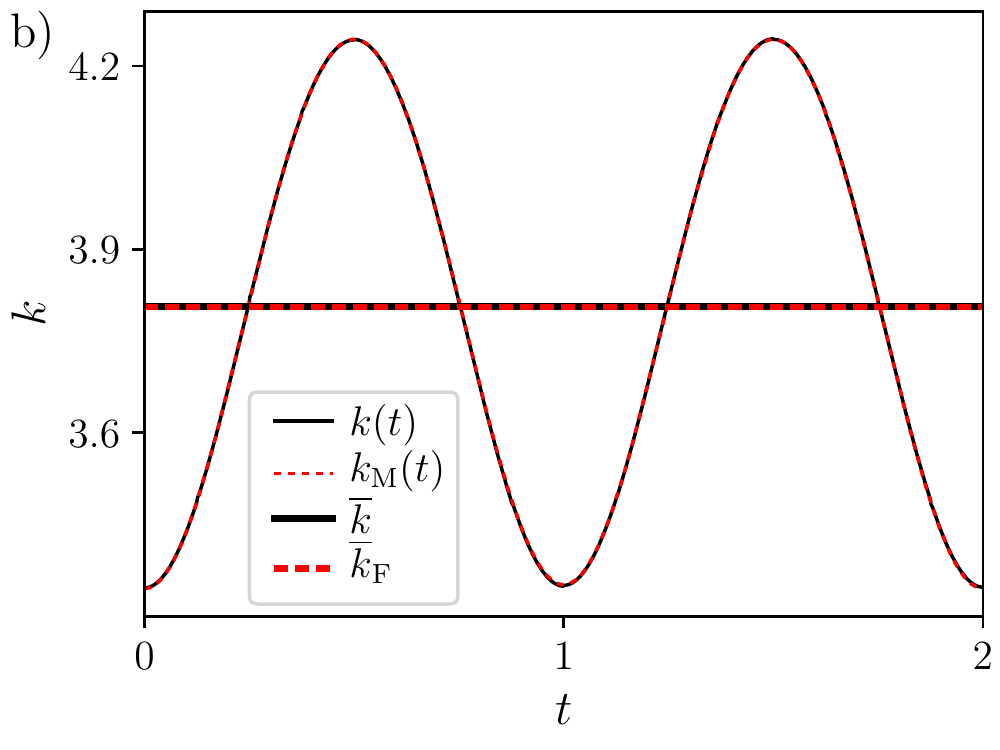}
  \caption{
	  (a) Reactant populations $N\sno{react}(t)$ when propagating
	  an ensemble of $200$ reactive trajectories.
	  The trajectories are initialized at $t_0=0$ with $\Delta x = 10^{-3}$
	  according to the method explained in Fig.~\ref{fig:ensemble_preparation}
	  close to the \ac{TS} trajectory
	  of the two-dimensional model system~\eqref{eq:potential}.
	  The straight lines yield $k(t)$ as instantaneous derivatives of $N\sno{react}(t)$,
	  taken at several times $t$ and visualized by crosses in the respective color.
	  (b)
	  Instantaneous ensemble rate $k(t)$ (thin black line),
	  obtained according to Sec.~\ref{sec:ensemble_rates} by numerical differentiation of
	  $N\sno{react}(t)$ for ensembles initialized close to the \ac{TS} trajectory
	  of the two-dimensional model system~\eqref{eq:potential} at different times $t_0$.
	  Here, $T=2$ is the period of the \ac{TS} trajectory.
	  The rate $k\sno{M}(t)$, calculated via the \ac{LMA}, as described in Sec.~\ref{sec:manifold_method},
	  is given as a thin, red dashed line, that coincides perfectly with $k(t)$,
	  hence yielding the same mean value $\overline{k}\sno{M} = \overline{k}$
	  when averaged over a full period of the oscillating barrier.
	  The Floquet rate, $\overline{k}\sno{F}=3.806$ (thick, red dashed line)
	  obtained by a linear regression of the red line in Fig.~\ref{fig:NHIM_cross_y_at_TSt}
	  is in perfect accordance to the mean values
	  $\overline{k}$ (thick black line) of both $k(t)$, as well as of $k\sno{M}(t)$.}
  \label{fig:test_system_TSt}
\end{figure}

We first employ the direct ensemble calculations along the \ac{TS}
trajectory of the two-dimensional model system~\eqref{eq:potential}
at different times
$t$, according to Sec.~\ref{sec:ensemble_preparation}.
A typical example of a reactant decay curve $N\sno{react}(t)$,
initialized with $200$ reacting trajectories
at $t_0 = 0$ with a distance $\Delta x = 10^{-3}$
is displayed in Fig.~\ref{fig:test_system_TSt}(a).
For several times $t$, visualized by the different colored crosses in
Fig.~\ref{fig:test_system_TSt}(a), the instantaneous
derivatives of $N\sno{react}(t)$ are shown as straight lines,
yielding the instantaneous rates $k(t)$ at the respective time.
As expected, the decay $N\sno{react}(t)$ is not purely exponential in the driven case.

The instantaneous rate $k(t)$, which is given as the
most accurate result from the numerical differentiation of the reactant decays
$N\sno{react}(t)$ of several ensembles (see Sec.~\ref{sec:ensemble_rate}),
is therefore not constant in time, as
seen in Fig.~\ref{fig:test_system_TSt}(b). For
this particular result, $16$ ensembles with $200$ trajectories have been
launched at equidistant times
$t \in [0,T]$, where $T$ is the oscillation period of the saddle
potential~\eqref{eq:potential}.

According to Fig.~\ref{fig:test_system_TSt}(b),
the instantaneous rate $k(t)$
of the \ac{TS} trajectory
shows an oscillation between $k\sno{min} \approx 3.4$
and $k\sno{max} \approx 4.3$.
The oscillation of $k(t)$ is twice as fast as
the oscillation of the driving in Eq.~\eqref{eq:potential}
with $T=2$
and, hence, of the \ac{TS} trajectory,

Introduced in Sec.~\ref{sec:manifold_method}, the \ac{LMA}
allows us to obtain instantaneous rates using just the geometrical
properties of phase space structures near a single trajectory on the \ac{NHIM},
without time-consuming propagation of reactive trajectory ensembles.
Based on Eq.~\eqref{eq:k_t_final}, the instantaneous manifold
rate $k\sno{M}(t)$ is shown as thin red dashed line in
Fig.~\ref{fig:test_system_TSt}(b). The perfect agreement between $k(t)$
and $k\sno{M}(t)$ confirms the reliability of the \ac{LMA}.

Based on a stability analysis of the \ac{TS} trajectory
summarized in Sec.~\ref{sec:floquet_method},
the Floquet rate $\overline{k}\sno{F}$ can also be obtained.
It is included in Fig.~\ref{fig:test_system_TSt}(b)
as a thick red dashed
horizontal line.
Since the Floquet rate $\overline{k}\sno{F}$ according to
Eqs.~\eqref{eq:floquet_rate} and \eqref{eq:floquet_exponents_periodic}
is an integrated quantity over the full
(periodic) \ac{TS} trajectory, it is only a single value,
valid for the full trajectory.
The instantaneous rates $k(t)$ and $k\sno{M}(t)$ seem to
oscillate around $\overline{k}\sno{F}$.
For a more detailed comparison, the mean value of
the instantaneous rates
$\overline{k} = \overline{k(t)} = \overline{k\sno{M}(t)}$
is included as thick black horizontal line in
Fig.~\ref{fig:test_system_TSt}(b).
The almost perfect agreement between $\overline{k} = 3.8067$
and $\overline{k}\sno{F} = 3.8064$
is clear.
This result shows that the additional information encoded
in the instantaneous reaction rate nevertheless
has a mean value that coincides with the
Floquet rate.
The latter is an integrated quantity for the stability of the
(periodic) trajectory which remains encoded in the integral
of the rates obtained by the other two methods.

\subsection{Rate calculation for arbitrary trajectories on the NHIM}
\label{sec:non-periodic_trajectories}

To explain the procedure for how to obtain reaction rates
associated with
non-periodic trajectories on the \ac{NHIM},
we investigate a quasi-periodic trajectory on a stable torus
around the \ac{TS} trajectory shown by the PSOS in
Fig.~\ref{fig:floquet_nhim}.
As in the periodic case reported in
Sec.~\ref{sec:periodic_trajectories}, an instantaneous rate
can be obtained by propagating several
ensembles of reactive trajectories that are initialized close to the
respective trajectory on the
stable torus.

\begin{figure}
  \includegraphics[width=0.95\columnwidth]{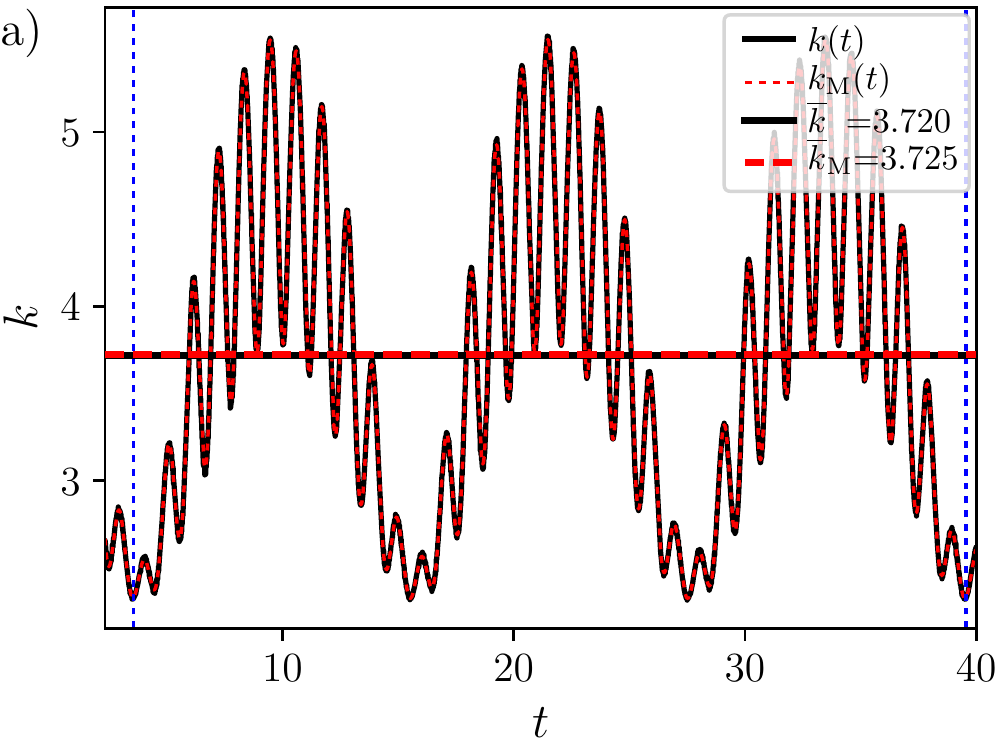}
  \includegraphics[width=0.95\columnwidth]{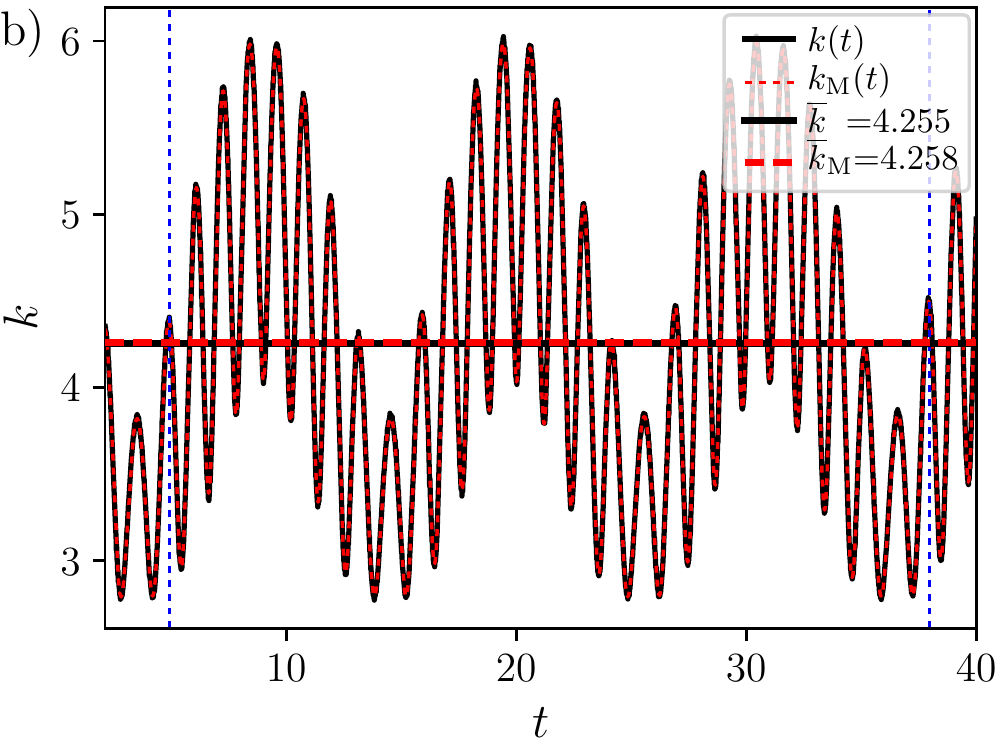}
  \caption{%
	  Instantaneous rate $k(t)$ (thin black line),
	  obtained using
	  a numerical differentiation of the reactant population $N\sno{react}(t)$
	  of $80$ ensembles with $200$ trajectories each, launched equidistantly in time
	  in between a time interval of $\Delta t = 40$.
	  All ensembles are prepared according to Sec.~\ref{sec:ensemble_preparation}
	  with $\Delta x = 10^{-3}$ close to a
	  trajectory that is initialized for $t=0$
	  at position
	  $y = 0.4$ and $v_y = -0.75$ (above) or $y = 0.7$ and $v_y = -0.75$ (below)
	  on the NHIM of the time-dependently
	  driven system defined in Eq.~\eqref{eq:potential}.
	  According to the procedure explained in Sec.~\ref{sec:manifold_method}
	  $k\sno{M}(t)$ is obtained and displayed
	  as thin, red dashed line, that perfectly matches $k(t)$.
	  The Floquet rate, calculated as explained in Sec.~\ref{sec:manifold_method}
	  or as displayed in Fig.~\ref{fig:NHIM_cross_y_at_TSt} is given as
	  thick, red dashed line. When averaging over three quasi-oscillations
	  of $k(t)$, visualized by the two vertical, blue dashed lines,
	  the Floquet rate perfectly matches
	  the mean ensemble rate $\overline{k}$ of both $k(t)$ as well as of $k\sno{M}(t)$.
  }
  \label{fig:test_system_torus_tr}
\end{figure}
In Fig.~\ref{fig:test_system_torus_tr}(a), the instantaneous
rate $k(t)$ is obtained by propagating $80$ ensembles
of $200$ trajectories.
Each ensemble is
initialized according to the procedure explained in Sec.~\ref{sec:ensemble_preparation}
with $\Delta x = 10^{-3}$ close
to the trajectory with initial conditions $t=0$ on the NHIM at position
$y = 0.4$ and $v_y = -0.75$.
The initial times $t_0 \in [0,40]$ of any of these
ensembles
are chosen in equally spaced intervals of $\Delta t = 0.5$.
Four ensembles are launched per period $T=2$ of the time-periodically driven
saddle according to Eq.~\eqref{eq:potential}.
As already explained in Secs.~\ref{sec:ensemble_rates} and
\ref{sec:periodic_trajectories}, the instantaneous
rate $k(t)$ is obtained by numerically deriving the reactant population
$N\sno{react}(t)$ of these reactive ensembles---cf.\ the
thin black line in
Fig.~\ref{fig:test_system_torus_tr}(a).

Whereas the instantaneous rate of the periodic
\ac{TS} trajectory in Fig.~\ref{fig:test_system_TSt}(b)
is itself periodic with the period of the external driving,
the instantaneous rate $k(t)$ of the non-periodic trajectory
in Fig.~\ref{fig:test_system_torus_tr}(a) clearly does not show such
behavior on the time-scale of $T=2$.
Indeed, there seems to be an approximate periodicity on a much larger time-scale
with three oscillations in between the two vertical, blue dashed lines.
The reason is given by the dynamics itself,
as the corresponding trajectory on the NHIM
is quasi-periodic, meaning it approximately orbits the stable torus three
times in this time interval.

Using the \ac{LMA} explained in Sec.~\ref{sec:manifold_method}
the manifold rate $k\sno{M}(t)$ is obtained---cf.\ the  thin red dashed line in
Fig.~\ref{fig:test_system_torus_tr}(a).
Just like in the periodic case of Sec.~\ref{sec:periodic_trajectories},
both $k(t)$ as well as $k\sno{M}(t)$ coincide perfectly.

\begin{figure}
  \includegraphics[width=0.95\columnwidth]{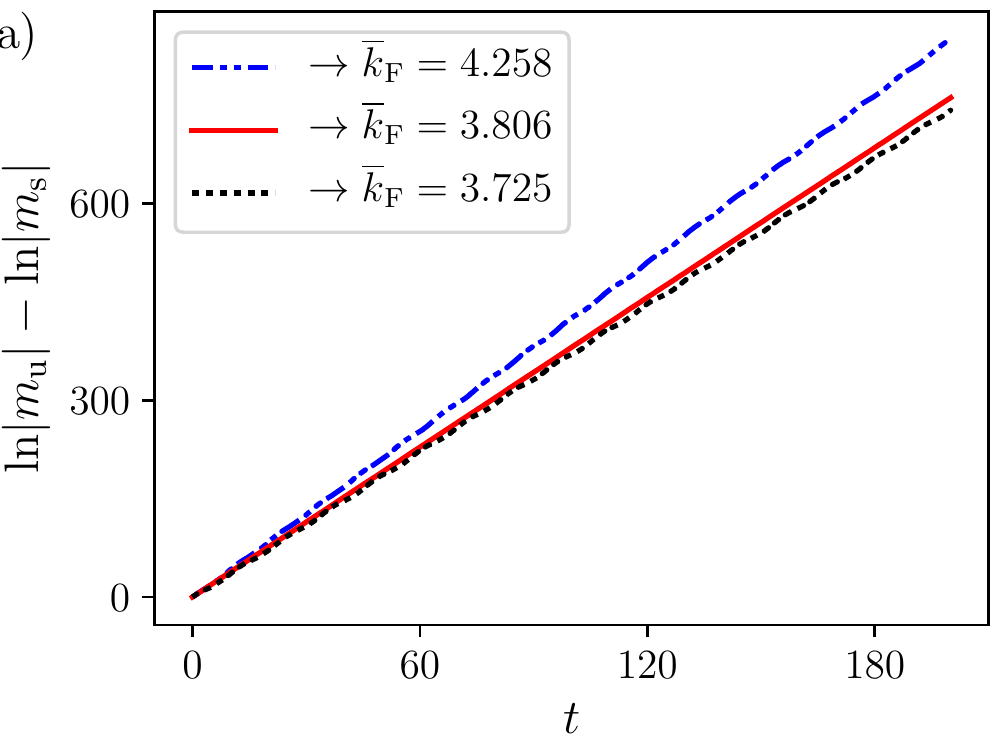}
  \includegraphics[width=0.95\columnwidth]{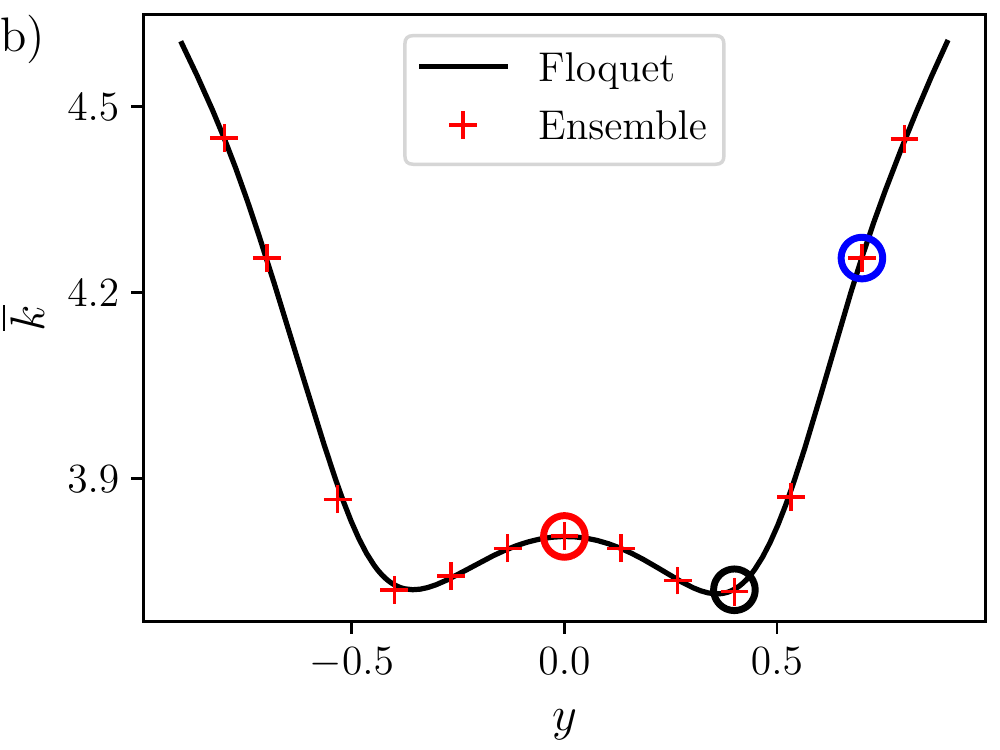}
  \caption{
	  (a): Difference $\ln |m\sno{u}| - \ln|m\sno{s}|$, where $m\sno{u,s}$ are the eigenvalues
	  of the time-dependent monodromy matrix along the unstable and stable directions
	  of the driven barrier according to Eq.~\eqref{eq:potential}.
	  The monodromy matrix is obtained time-dependently according to Eq.~\eqref{eq:mdot}
	  for the periodic TS trajectory
	  of Fig.~\ref{fig:test_system_TSt} (red line), as well as
	  for the two non-periodic trajectories introduced in Fig.~\ref{fig:test_system_torus_tr}
	  (black dotted line and blue dash-dotted line). The Floquet rates $\overline{k}\sno{F}$
	  in the limit $t \rightarrow \infty$ are obtained
	  according to Eq.~\eqref{eq:floquet_rate_non_periodic}
	  by a linear regression to these curves.
	  (b): Black line: calculated Floquet rates for trajectories initialized on
	  a cross-section at the velocity $v_{y}$ of the TS trajectory
	  of the \ac{NHIM}
	  at time $t = 0$. This corresponds
	  to a horizontal line through the center of
	  the tori in Fig.~\ref{fig:floquet_nhim}. To verify these results,
	  the mean ensemble rates (red crosses)
	  are obtained according to Sec.~\ref{sec:ensemble_rates}
	  for trajectories initialized at several values $y$ on this cross section.
	  The colored circles mark the corresponding Floquet calculations
	  of Fig.~\ref{fig:NHIM_cross_y_at_TSt}(a).
  }
  \label{fig:NHIM_cross_y_at_TSt}
\end{figure}

As introduced in Sec.~\ref{sec:floquet_method}, according to
Eq.~\eqref{eq:floquet_rate_non_periodic} a Floquet rate can be
obtained from the stability of a non-periodic trajectory.
As infinite time
integration cannot be achieved numerically, one has to pay special attention
to the convergence of the Floquet rate.
The time-dependent logarithmic
differences of the eigenvalues of the monodromy matrix
are shown in Fig.~\ref{fig:NHIM_cross_y_at_TSt}(a).
They are obtained according to Eq.~\eqref{eq:floquet_rate_non_periodic},
and shown as the black dotted curve
for an integration along the same trajectory
as in Fig.~\ref{fig:test_system_torus_tr}(a).
Here, a very long integration time
(up to $t=200$ in dimensionless units and corresponding to
100 quasi-periods) is chosen to ensure convergence.
The convergence value is extrapolated using linear regression and
a Floquet rate of
$\overline{k}\sno{F} = 3.725$ is obtained as the slope of the data
shown in Fig.~\ref{fig:NHIM_cross_y_at_TSt}(a).
The Floquet rate of the
trajectory is also shown
as a thick red dashed horizontal line in
Fig.~\ref{fig:test_system_torus_tr}(a),
and the agreement is remarkably good.

Averaging over the three quasi-oscillations
of $k(t)$ or $k\sno{M}(t)$, indicated by the blue dashed
vertical lines in Fig.~\ref{fig:test_system_torus_tr}(a)
yields the average rate $\overline{k} = 3.720$,
included as a thick black horizontal line in
Fig.~\ref{fig:test_system_torus_tr}(a),
that again is in quite good agreement
with the Floquet rate $\overline{k}\sno{F} = 3.725$ obtained by
the stability analysis of the underlying trajectory on the stable torus.
Hence, even for non-periodic trajectories,
the mean value of the instantaneous rates corresponds
to the Floquet rate obtained according to
Eq.~\eqref{eq:floquet_rate_non_periodic}.

This provides a hint to another important result.
The mean rate obtained for the
periodic \ac{TS} trajectory in Fig.~\ref{fig:test_system_TSt}
is different
than the rate for the non-periodic trajectory
of Fig.~\ref{fig:test_system_torus_tr}(a)
on a stable torus around the \ac{TS} trajectory.
This implies that the rate
depends on the position of a trajectory on the NHIM.
To test this, an alternative calculation
not-quite-similar to the one of Fig.~\ref{fig:test_system_torus_tr}(a)
was performed on a trajectory that starts for $t=0$
at position $y=0.7$ and velocity $v_y=-0.75$
further away from the \ac{TS} trajectory on the NHIM.
The results obtained with the same procedure as in
Fig.~\ref{fig:test_system_torus_tr}(a)
are shown in Fig.~\ref{fig:test_system_torus_tr}(b)
Again, not only is the instantaneous rate $k(t)$
obtained by propagating an ensemble
of reactive trajectories
in near perfect agreement with
the manifold rate $k\sno{M}(t)$, but also the
accordance between their average rate $\overline{k} = 4.255$
and the Floquet rate $\overline{k}\sno{F} = 4.258$, obtained by linear
regression of the blue dash-dotted curve in
Fig.~\ref{fig:NHIM_cross_y_at_TSt}(a) is
pretty good.
And again, the mean rates obtained here are different
to the mean rates of the periodic \ac{TS} trajectory
in Fig.~\ref{fig:test_system_TSt} or the non-periodic
torus trajectory of Fig.~\ref{fig:test_system_torus_tr}(a).

To investigate the trajectory dependence of the rates further,
we employed the Floquet method
to calculate the mean rates for many trajectories on the \ac{NHIM}
initialized in the intervals $y \in [-0.8, 0.8]$ and
$v_y \in [-2, 2]$ at time $t = 0$.
The results are given as color encoded as per
Fig.~\ref{fig:floquet_nhim}. Note,
that although we show Floquet rates here,
we could just as well use the mean rates obtained from one
of the instantaneous rate methods, because all of these methods
lead to the same average rates.

For the quasi-periodic dynamics on the tori in
Fig.~\ref{fig:floquet_nhim}, any torus is associated to its
own mean rate.
In an area around the \ac{TS} trajectory, we observe that the
mean rates first decrease, before increasing significantly.
Taking a closer look, the rate of the \ac{TS} trajectory
of the model system~\eqref{eq:potential}
is a local maximum in the global minimum of reaction
rates on the NHIM.
This can be seen better
by using a $y$ cross section through Fig.~\ref{fig:floquet_nhim}
at the velocity $v_y$ of the \ac{TS} trajectory,
see Fig.~\ref{fig:NHIM_cross_y_at_TSt}(b).
Here, we support the Floquet calculations (black line)
with the average rates obtained by the more time-consuming
ensemble propagations (red crosses).
We also highlight the \ac{TS} trajectory regarded in
Fig.~\ref{fig:test_system_TSt} with a red circle,
and the two trajectories of Fig.~\ref{fig:test_system_torus_tr}
with a black and a blue circle.

\section{Conclusion and Outlook}
\label{sec:Conclusion}
In this paper, we have shown how to obtain
phase-space resolved rates in driven multidimensional
chemical reactions.
Three methods are compared.
The first is based on measuring the flux of trajectories
through a recrossing-free \ac{DS} attached to the
time-dependently moving \ac{NHIM} in the barrier region
of a rank-1 saddle separating reactants from products.
For the second \ac{LMA} method,
rates are obtained directly using just the geometrical
properties of phase space close to a single trajectory
located within the \ac{NHIM} of the saddle.
No time-consuming trajectory propagation is needed here.
The third method is a multidimensional generalization of the Floquet method
that has already been applied to obtain the rates of a
one-dimensional system~\cite{hern14f}.
We provide evidence that the first two methods lead to the same result
for the instantaneous reaction rate.
Their mean (or long time) average corresponds perfectly
to the generalized Floquet rates that are based on the stability
analysis of a single trajectory.

Whereas the rates in this paper are the decay rates
of trajectories contained in
the activated complex near the \ac{TS} in between
reactants and products, the rates measured
in real and thermally activated systems correspond to
the Kramers rate. Finding a connection between
these two different rates
can be the subject of future work.
  So far, the instantaneous decay rates of trajectories
  in the \ac{TS} can be determined exactly using the \ac{LMA}
  and these decay rates are also valid in a certain area
  close to the \ac{TS} as has been shown via the ensemble method.
  In the nearby neighborhood of the \ac{NHIM},
  reactive trajectories correspond to the slowest
  flux and, hence, are critical in determining
  the rate of reactions
  through a time-dependent \ac{DS}.
  Since this \ac{DS} is attached to the \ac{NHIM}
  it is, at least in that nearby neighborhood, free of recrossings.
  The question on how large this ``nearby'' region needs to be
  to yield conclusive results is also subject to
  further work, and likely depends on the specific system.

It would also be interesting to determine
how reactions can be controlled
and what effects can be seen when changing the parameters of
the external driving.
A different challenge for future research is the application
of the methods developed here to
specific model chemical reactions.
For example, the rank-1 barrier in between the two
(meta)-stable conformations in the LiCN $\leftrightarrow$ LiNC
isomerization reaction may be an easy example to test
the methods presented here.
An application to KCN~\cite{borondo18a} or Ketene~\cite{hern16d} is also conceivable,
but since these systems contain more than one barrier
in between their reactant and product states,
the dynamics is expected to be much more complex and chaotic.

\section*{Acknowledgments}
\label{sec:Acknowledgments}
The German portion of this collaborative work was supported
by Deutsche Forschungsgemeinschaft (DFG) through Grant
No.~MA1639/14-1.
RH's contribution to this work was supported by the National Science
Foundation (NSF) through Grant No.~CHE-1700749.
M.F.\ is grateful for support from the Landesgraduiertenf\"orderung of
the Land Baden-W\"urttemberg.
This collaboration has also benefited from support by the European
Union's Horizon 2020 Research and Innovation Program under the Marie
Sklodowska-Curie Grant Agreement No.~734557.

\section*{References}
\bibliography{paperq25long}

\begin{thebibliography}{45}%
\makeatletter
\providecommand \@ifxundefined [1]{%
 \@ifx{#1\undefined}
}%
\providecommand \@ifnum [1]{%
 \ifnum #1\expandafter \@firstoftwo
 \else \expandafter \@secondoftwo
 \fi
}%
\providecommand \@ifx [1]{%
 \ifx #1\expandafter \@firstoftwo
 \else \expandafter \@secondoftwo
 \fi
}%
\providecommand \natexlab [1]{#1}%
\providecommand \enquote  [1]{``#1''}%
\providecommand \bibnamefont  [1]{#1}%
\providecommand \bibfnamefont [1]{#1}%
\providecommand \citenamefont [1]{#1}%
\providecommand \href@noop [0]{\@secondoftwo}%
\providecommand \href [0]{\begingroup \@sanitize@url \@href}%
\providecommand \@href[1]{\@@startlink{#1}\@@href}%
\providecommand \@@href[1]{\endgroup#1\@@endlink}%
\providecommand \@sanitize@url [0]{\catcode `\\12\catcode `\$12\catcode
  `\&12\catcode `\#12\catcode `\^12\catcode `\_12\catcode `\%12\relax}%
\providecommand \@@startlink[1]{}%
\providecommand \@@endlink[0]{}%
\providecommand \url  [0]{\begingroup\@sanitize@url \@url }%
\providecommand \@url [1]{\endgroup\@href {#1}{\urlprefix }}%
\providecommand \urlprefix  [0]{URL }%
\providecommand \Eprint [0]{\href }%
\providecommand \doibase [0]{http://dx.doi.org/}%
\providecommand \selectlanguage [0]{\@gobble}%
\providecommand \bibinfo  [0]{\@secondoftwo}%
\providecommand \bibfield  [0]{\@secondoftwo}%
\providecommand \translation [1]{[#1]}%
\providecommand \BibitemOpen [0]{}%
\providecommand \bibitemStop [0]{}%
\providecommand \bibitemNoStop [0]{.\EOS\space}%
\providecommand \EOS [0]{\spacefactor3000\relax}%
\providecommand \BibitemShut  [1]{\csname bibitem#1\endcsname}%
\let\auto@bib@innerbib\@empty
\bibitem [{\citenamefont {Eyring}(1935)}]{eyring35}%
  \BibitemOpen
  \bibfield  {author} {\bibinfo {author} {\bibfnamefont {H.}~\bibnamefont
  {Eyring}},\ }\href@noop {} {\bibfield  {journal} {\bibinfo  {journal} {J.
  Chem. Phys.}\ }\textbf {\bibinfo {volume} {3}},\ \bibinfo {pages} {107}
  (\bibinfo {year} {1935})}\BibitemShut {NoStop}%
\bibitem [{\citenamefont {Wigner}(1937)}]{wigner37}%
  \BibitemOpen
  \bibfield  {author} {\bibinfo {author} {\bibfnamefont {E.~P.}\ \bibnamefont
  {Wigner}},\ }\href@noop {} {\bibfield  {journal} {\bibinfo  {journal} {J.
  Chem. Phys.}\ }\textbf {\bibinfo {volume} {5}},\ \bibinfo {pages} {720}
  (\bibinfo {year} {1937})}\BibitemShut {NoStop}%
\bibitem [{\citenamefont {Pechukas}(1981)}]{pech81}%
  \BibitemOpen
  \bibfield  {author} {\bibinfo {author} {\bibfnamefont {P.}~\bibnamefont
  {Pechukas}},\ }\href@noop {} {\bibfield  {journal} {\bibinfo  {journal}
  {Annu. Rev. Phys. Chem.}\ }\textbf {\bibinfo {volume} {32}},\ \bibinfo
  {pages} {159} (\bibinfo {year} {1981})}\BibitemShut {NoStop}%
\bibitem [{\citenamefont {Truhlar}, \citenamefont {Garrett},\ and\
  \citenamefont {Klippenstein}(1996)}]{truh96}%
  \BibitemOpen
  \bibfield  {author} {\bibinfo {author} {\bibfnamefont {D.~G.}\ \bibnamefont
  {Truhlar}}, \bibinfo {author} {\bibfnamefont {B.~C.}\ \bibnamefont
  {Garrett}}, \ and\ \bibinfo {author} {\bibfnamefont {S.~J.}\ \bibnamefont
  {Klippenstein}},\ }\href@noop {} {\bibfield  {journal} {\bibinfo  {journal}
  {J. Phys. Chem.}\ }\textbf {\bibinfo {volume} {100}},\ \bibinfo {pages}
  {12771} (\bibinfo {year} {1996})}\BibitemShut {NoStop}%
\bibitem [{\citenamefont {Mullen}, \citenamefont {Shea},\ and\ \citenamefont
  {Peters}(2014)}]{peters14a}%
  \BibitemOpen
  \bibfield  {author} {\bibinfo {author} {\bibfnamefont {R.~G.}\ \bibnamefont
  {Mullen}}, \bibinfo {author} {\bibfnamefont {J.-E.}\ \bibnamefont {Shea}}, \
  and\ \bibinfo {author} {\bibfnamefont {B.}~\bibnamefont {Peters}},\
  }\href@noop {} {\bibfield  {journal} {\bibinfo  {journal} {J. Chem. Phys.}\
  }\textbf {\bibinfo {volume} {140}},\ \bibinfo {pages} {041104} (\bibinfo
  {year} {2014})}\BibitemShut {NoStop}%
\bibitem [{\citenamefont {Wiggins}(2016)}]{wiggins16}%
  \BibitemOpen
  \bibfield  {author} {\bibinfo {author} {\bibfnamefont {S.}~\bibnamefont
  {Wiggins}},\ }\href@noop {} {\bibfield  {journal} {\bibinfo  {journal}
  {Regul. Chaotic Dyn.}\ }\textbf {\bibinfo {volume} {21}},\ \bibinfo {pages}
  {621} (\bibinfo {year} {2016})}\BibitemShut {NoStop}%
\bibitem [{\citenamefont {Feldmaier}\ \emph {et~al.}(2019)\citenamefont
  {Feldmaier}, \citenamefont {Schraft}, \citenamefont {Bardakcioglu},
  \citenamefont {Reiff}, \citenamefont {Lober}, \citenamefont {Tsch{\"o}pe},
  \citenamefont {Junginger}, \citenamefont {Main}, \citenamefont {Bartsch},\
  and\ \citenamefont {Hernandez}}]{hern19a}%
  \BibitemOpen
  \bibfield  {author} {\bibinfo {author} {\bibfnamefont {M.}~\bibnamefont
  {Feldmaier}}, \bibinfo {author} {\bibfnamefont {P.}~\bibnamefont {Schraft}},
  \bibinfo {author} {\bibfnamefont {R.}~\bibnamefont {Bardakcioglu}}, \bibinfo
  {author} {\bibfnamefont {J.}~\bibnamefont {Reiff}}, \bibinfo {author}
  {\bibfnamefont {M.}~\bibnamefont {Lober}}, \bibinfo {author} {\bibfnamefont
  {M.}~\bibnamefont {Tsch{\"o}pe}}, \bibinfo {author} {\bibfnamefont
  {A.}~\bibnamefont {Junginger}}, \bibinfo {author} {\bibfnamefont
  {J.}~\bibnamefont {Main}}, \bibinfo {author} {\bibfnamefont {T.}~\bibnamefont
  {Bartsch}}, \ and\ \bibinfo {author} {\bibfnamefont {R.}~\bibnamefont
  {Hernandez}},\ }\href@noop {} {\bibfield  {journal} {\bibinfo  {journal} {J.
  Phys. Chem. B}\ }\textbf {\bibinfo {volume} {123}},\ \bibinfo {pages} {2070}
  (\bibinfo {year} {2019})}\BibitemShut {NoStop}%
\bibitem [{\citenamefont {Lorquet}(2017)}]{lorquet17}%
  \BibitemOpen
  \bibfield  {author} {\bibinfo {author} {\bibfnamefont {J.~C.}\ \bibnamefont
  {Lorquet}},\ }\href@noop {} {\bibfield  {journal} {\bibinfo  {journal} {J.
  Chem. Phys.}\ }\textbf {\bibinfo {volume} {146}},\ \bibinfo {pages} {134310}
  (\bibinfo {year} {2017})}\BibitemShut {NoStop}%
\bibitem [{\citenamefont {Patra}\ and\ \citenamefont
  {Keshavamurthy}(2018)}]{keshavamurthy18}%
  \BibitemOpen
  \bibfield  {author} {\bibinfo {author} {\bibfnamefont {S.}~\bibnamefont
  {Patra}}\ and\ \bibinfo {author} {\bibfnamefont {S.}~\bibnamefont
  {Keshavamurthy}},\ }\href@noop {} {\bibfield  {journal} {\bibinfo  {journal}
  {Phys. Chem. Chem. Phys.}\ }\textbf {\bibinfo {volume} {20}},\ \bibinfo
  {pages} {4970} (\bibinfo {year} {2018})}\BibitemShut {NoStop}%
\bibitem [{\citenamefont {Kraj{\u{n}}{\'{a}}k}\ and\ \citenamefont
  {Waalkens}(2018)}]{waalkens18}%
  \BibitemOpen
  \bibfield  {author} {\bibinfo {author} {\bibfnamefont {V.}~\bibnamefont
  {Kraj{\u{n}}{\'{a}}k}}\ and\ \bibinfo {author} {\bibfnamefont
  {H.}~\bibnamefont {Waalkens}},\ }\href@noop {} {\bibfield  {journal}
  {\bibinfo  {journal} {J. Math. Chem.}\ }\textbf {\bibinfo {volume} {56}},\
  \bibinfo {pages} {2341—2378} (\bibinfo {year} {2018})}\BibitemShut
  {NoStop}%
\bibitem [{\citenamefont {Tamiya}\ \emph {et~al.}(2018)\citenamefont {Tamiya},
  \citenamefont {Watanabe}, \citenamefont {Noji}, \citenamefont {Li},\ and\
  \citenamefont {Komatsuzaki}}]{komatsuzaki18}%
  \BibitemOpen
  \bibfield  {author} {\bibinfo {author} {\bibfnamefont {Y.}~\bibnamefont
  {Tamiya}}, \bibinfo {author} {\bibfnamefont {R.}~\bibnamefont {Watanabe}},
  \bibinfo {author} {\bibfnamefont {H.}~\bibnamefont {Noji}}, \bibinfo {author}
  {\bibfnamefont {C.-B.}\ \bibnamefont {Li}}, \ and\ \bibinfo {author}
  {\bibfnamefont {T.}~\bibnamefont {Komatsuzaki}},\ }\href@noop {} {\bibfield
  {journal} {\bibinfo  {journal} {Phys. Chem. Chem. Phys.}\ }\textbf {\bibinfo
  {volume} {20}},\ \bibinfo {pages} {1872} (\bibinfo {year}
  {2018})}\BibitemShut {NoStop}%
\bibitem [{\citenamefont {Hernandez}\ and\ \citenamefont
  {Miller}(1993)}]{hern93b}%
  \BibitemOpen
  \bibfield  {author} {\bibinfo {author} {\bibfnamefont {R.}~\bibnamefont
  {Hernandez}}\ and\ \bibinfo {author} {\bibfnamefont {W.~H.}\ \bibnamefont
  {Miller}},\ }\href@noop {} {\bibfield  {journal} {\bibinfo  {journal} {Chem.
  Phys. Lett.}\ }\textbf {\bibinfo {volume} {214}},\ \bibinfo {pages} {129}
  (\bibinfo {year} {1993})}\BibitemShut {NoStop}%
\bibitem [{\citenamefont {Wiggins}\ \emph {et~al.}(2001)\citenamefont
  {Wiggins}, \citenamefont {Wiesenfeld}, \citenamefont {Jaffe},\ and\
  \citenamefont {Uzer}}]{wiggins01}%
  \BibitemOpen
  \bibfield  {author} {\bibinfo {author} {\bibfnamefont {S.}~\bibnamefont
  {Wiggins}}, \bibinfo {author} {\bibfnamefont {L.}~\bibnamefont {Wiesenfeld}},
  \bibinfo {author} {\bibfnamefont {C.}~\bibnamefont {Jaffe}}, \ and\ \bibinfo
  {author} {\bibfnamefont {T.}~\bibnamefont {Uzer}},\ }\href@noop {} {\bibfield
   {journal} {\bibinfo  {journal} {Phys. Rev. Lett.}\ }\textbf {\bibinfo
  {volume} {86}} (\bibinfo {year} {2001})}\BibitemShut {NoStop}%
\bibitem [{\citenamefont {Uzer}\ \emph {et~al.}(2002)\citenamefont {Uzer},
  \citenamefont {Jaff\'e}, \citenamefont {Palaci{\'a}n}, \citenamefont
  {Yanguas},\ and\ \citenamefont {Wiggins}}]{Uzer02}%
  \BibitemOpen
  \bibfield  {author} {\bibinfo {author} {\bibfnamefont {T.}~\bibnamefont
  {Uzer}}, \bibinfo {author} {\bibfnamefont {C.}~\bibnamefont {Jaff\'e}},
  \bibinfo {author} {\bibfnamefont {J.}~\bibnamefont {Palaci{\'a}n}}, \bibinfo
  {author} {\bibfnamefont {P.}~\bibnamefont {Yanguas}}, \ and\ \bibinfo
  {author} {\bibfnamefont {S.}~\bibnamefont {Wiggins}},\ }\href@noop {}
  {\bibfield  {journal} {\bibinfo  {journal} {Nonlinearity}\ }\textbf {\bibinfo
  {volume} {15}},\ \bibinfo {pages} {957} (\bibinfo {year} {2002})}\BibitemShut
  {NoStop}%
\bibitem [{\citenamefont {Bartsch}, \citenamefont {Hernandez},\ and\
  \citenamefont {Uzer}(2005)}]{dawn05a}%
  \BibitemOpen
  \bibfield  {author} {\bibinfo {author} {\bibfnamefont {T.}~\bibnamefont
  {Bartsch}}, \bibinfo {author} {\bibfnamefont {R.}~\bibnamefont {Hernandez}},
  \ and\ \bibinfo {author} {\bibfnamefont {T.}~\bibnamefont {Uzer}},\
  }\href@noop {} {\bibfield  {journal} {\bibinfo  {journal} {Phys. Rev. Lett.}\
  }\textbf {\bibinfo {volume} {95}},\ \bibinfo {pages} {058301} (\bibinfo
  {year} {2005})}\BibitemShut {NoStop}%
\bibitem [{\citenamefont {Bartsch}\ \emph {et~al.}(2008)\citenamefont
  {Bartsch}, \citenamefont {Moix}, \citenamefont {Hernandez}, \citenamefont
  {Kawai},\ and\ \citenamefont {Uzer}}]{hern08d}%
  \BibitemOpen
  \bibfield  {author} {\bibinfo {author} {\bibfnamefont {T.}~\bibnamefont
  {Bartsch}}, \bibinfo {author} {\bibfnamefont {J.~M.}\ \bibnamefont {Moix}},
  \bibinfo {author} {\bibfnamefont {R.}~\bibnamefont {Hernandez}}, \bibinfo
  {author} {\bibfnamefont {S.}~\bibnamefont {Kawai}}, \ and\ \bibinfo {author}
  {\bibfnamefont {T.}~\bibnamefont {Uzer}},\ }\href@noop {} {\bibfield
  {journal} {\bibinfo  {journal} {Adv. Chem. Phys.}\ }\textbf {\bibinfo
  {volume} {140}},\ \bibinfo {pages} {191} (\bibinfo {year}
  {2008})}\BibitemShut {NoStop}%
\bibitem [{\citenamefont {Hernandez}(1994)}]{hern94}%
  \BibitemOpen
  \bibfield  {author} {\bibinfo {author} {\bibfnamefont {R.}~\bibnamefont
  {Hernandez}},\ }\href@noop {} {\bibfield  {journal} {\bibinfo  {journal} {J.
  Chem. Phys.}\ }\textbf {\bibinfo {volume} {101}},\ \bibinfo {pages} {9534}
  (\bibinfo {year} {1994})}\BibitemShut {NoStop}%
\bibitem [{\citenamefont {Allahem}\ and\ \citenamefont
  {Bartsch}(2012)}]{Allahem12}%
  \BibitemOpen
  \bibfield  {author} {\bibinfo {author} {\bibfnamefont {A.}~\bibnamefont
  {Allahem}}\ and\ \bibinfo {author} {\bibfnamefont {T.}~\bibnamefont
  {Bartsch}},\ }\href@noop {} {\bibfield  {journal} {\bibinfo  {journal} {J.
  Chem. Phys.}\ }\textbf {\bibinfo {volume} {137}},\ \bibinfo {pages} {214310}
  (\bibinfo {year} {2012})}\BibitemShut {NoStop}%
\bibitem [{\citenamefont {Junginger}\ \emph {et~al.}(2016)\citenamefont
  {Junginger}, \citenamefont {Garcia-Muller}, \citenamefont {Borondo},
  \citenamefont {Benito},\ and\ \citenamefont {Hernandez}}]{hern16c}%
  \BibitemOpen
  \bibfield  {author} {\bibinfo {author} {\bibfnamefont {A.}~\bibnamefont
  {Junginger}}, \bibinfo {author} {\bibfnamefont {P.~L.}\ \bibnamefont
  {Garcia-Muller}}, \bibinfo {author} {\bibfnamefont {F.}~\bibnamefont
  {Borondo}}, \bibinfo {author} {\bibfnamefont {R.~M.}\ \bibnamefont {Benito}},
  \ and\ \bibinfo {author} {\bibfnamefont {R.}~\bibnamefont {Hernandez}},\
  }\href@noop {} {\bibfield  {journal} {\bibinfo  {journal} {J. Chem. Phys.}\
  }\textbf {\bibinfo {volume} {144}},\ \bibinfo {pages} {024104} (\bibinfo
  {year} {2016})}\BibitemShut {NoStop}%
\bibitem [{\citenamefont {Craven}\ and\ \citenamefont
  {Hernandez}(2016)}]{hern16d}%
  \BibitemOpen
  \bibfield  {author} {\bibinfo {author} {\bibfnamefont {G.~T.}\ \bibnamefont
  {Craven}}\ and\ \bibinfo {author} {\bibfnamefont {R.}~\bibnamefont
  {Hernandez}},\ }\href@noop {} {\bibfield  {journal} {\bibinfo  {journal}
  {Phys. Chem. Chem. Phys.}\ }\textbf {\bibinfo {volume} {18}},\ \bibinfo
  {pages} {4008} (\bibinfo {year} {2016})}\BibitemShut {NoStop}%
\bibitem [{\citenamefont {Feldmaier}\ \emph {et~al.}(2017)\citenamefont
  {Feldmaier}, \citenamefont {Junginger}, \citenamefont {Main}, \citenamefont
  {Wunner},\ and\ \citenamefont {Hernandez}}]{hern17h}%
  \BibitemOpen
  \bibfield  {author} {\bibinfo {author} {\bibfnamefont {M.}~\bibnamefont
  {Feldmaier}}, \bibinfo {author} {\bibfnamefont {A.}~\bibnamefont
  {Junginger}}, \bibinfo {author} {\bibfnamefont {J.}~\bibnamefont {Main}},
  \bibinfo {author} {\bibfnamefont {G.}~\bibnamefont {Wunner}}, \ and\ \bibinfo
  {author} {\bibfnamefont {R.}~\bibnamefont {Hernandez}},\ }\href@noop {}
  {\bibfield  {journal} {\bibinfo  {journal} {Chem. Phys. Lett.}\ }\textbf
  {\bibinfo {volume} {687}},\ \bibinfo {pages} {194} (\bibinfo {year}
  {2017})}\BibitemShut {NoStop}%
\bibitem [{\citenamefont {Pollak}\ and\ \citenamefont
  {Pechukas}(1978)}]{pollak78}%
  \BibitemOpen
  \bibfield  {author} {\bibinfo {author} {\bibfnamefont {E.}~\bibnamefont
  {Pollak}}\ and\ \bibinfo {author} {\bibfnamefont {P.}~\bibnamefont
  {Pechukas}},\ }\href@noop {} {\bibfield  {journal} {\bibinfo  {journal} {J.
  Chem. Phys.}\ }\textbf {\bibinfo {volume} {69}},\ \bibinfo {pages} {1218}
  (\bibinfo {year} {1978})}\BibitemShut {NoStop}%
\bibitem [{\citenamefont {Pechukas}\ and\ \citenamefont
  {Pollak}(1979)}]{pech79a}%
  \BibitemOpen
  \bibfield  {author} {\bibinfo {author} {\bibfnamefont {P.}~\bibnamefont
  {Pechukas}}\ and\ \bibinfo {author} {\bibfnamefont {E.}~\bibnamefont
  {Pollak}},\ }\href@noop {} {\bibfield  {journal} {\bibinfo  {journal} {J.
  Chem. Phys.}\ }\textbf {\bibinfo {volume} {71}},\ \bibinfo {pages} {2062}
  (\bibinfo {year} {1979})}\BibitemShut {NoStop}%
\bibitem [{\citenamefont {Jaff{\'e}}\ \emph {et~al.}(2002)\citenamefont
  {Jaff{\'e}}, \citenamefont {Ross}, \citenamefont {Lo}, \citenamefont
  {Marsden}, \citenamefont {Farrelly},\ and\ \citenamefont {Uzer}}]{Jaffe02}%
  \BibitemOpen
  \bibfield  {author} {\bibinfo {author} {\bibfnamefont {C.}~\bibnamefont
  {Jaff{\'e}}}, \bibinfo {author} {\bibfnamefont {S.~D.}\ \bibnamefont {Ross}},
  \bibinfo {author} {\bibfnamefont {M.~W.}\ \bibnamefont {Lo}}, \bibinfo
  {author} {\bibfnamefont {J.}~\bibnamefont {Marsden}}, \bibinfo {author}
  {\bibfnamefont {D.}~\bibnamefont {Farrelly}}, \ and\ \bibinfo {author}
  {\bibfnamefont {T.}~\bibnamefont {Uzer}},\ }\href@noop {} {\bibfield
  {journal} {\bibinfo  {journal} {Phys. Rev. Lett.}\ }\textbf {\bibinfo
  {volume} {89}},\ \bibinfo {pages} {011101} (\bibinfo {year}
  {2002})}\BibitemShut {NoStop}%
\bibitem [{\citenamefont {Teramoto}, \citenamefont {Toda},\ and\ \citenamefont
  {Komatsuzaki}(2011)}]{Teramoto11}%
  \BibitemOpen
  \bibfield  {author} {\bibinfo {author} {\bibfnamefont {H.}~\bibnamefont
  {Teramoto}}, \bibinfo {author} {\bibfnamefont {M.}~\bibnamefont {Toda}}, \
  and\ \bibinfo {author} {\bibfnamefont {T.}~\bibnamefont {Komatsuzaki}},\
  }\href@noop {} {\bibfield  {journal} {\bibinfo  {journal} {Phys. Rev. Lett.}\
  }\textbf {\bibinfo {volume} {106}},\ \bibinfo {pages} {054101} (\bibinfo
  {year} {2011})}\BibitemShut {NoStop}%
\bibitem [{\citenamefont {Li}\ \emph {et~al.}(2006)\citenamefont {Li},
  \citenamefont {Shoujiguchi}, \citenamefont {Toda},\ and\ \citenamefont
  {Komatsuzaki}}]{komatsuzaki06a}%
  \BibitemOpen
  \bibfield  {author} {\bibinfo {author} {\bibfnamefont {C.-B.}\ \bibnamefont
  {Li}}, \bibinfo {author} {\bibfnamefont {A.}~\bibnamefont {Shoujiguchi}},
  \bibinfo {author} {\bibfnamefont {M.}~\bibnamefont {Toda}}, \ and\ \bibinfo
  {author} {\bibfnamefont {T.}~\bibnamefont {Komatsuzaki}},\ }\href@noop {}
  {\bibfield  {journal} {\bibinfo  {journal} {Phys. Rev. Lett.}\ }\textbf
  {\bibinfo {volume} {97}},\ \bibinfo {pages} {028302} (\bibinfo {year}
  {2006})}\BibitemShut {NoStop}%
\bibitem [{\citenamefont {Waalkens}\ and\ \citenamefont
  {Wiggins}(2004)}]{Waalkens04b}%
  \BibitemOpen
  \bibfield  {author} {\bibinfo {author} {\bibfnamefont {H.}~\bibnamefont
  {Waalkens}}\ and\ \bibinfo {author} {\bibfnamefont {S.}~\bibnamefont
  {Wiggins}},\ }\href@noop {} {\bibfield  {journal} {\bibinfo  {journal} {J.
  Phys. A}\ }\textbf {\bibinfo {volume} {37}},\ \bibinfo {pages} {L435}
  (\bibinfo {year} {2004})}\BibitemShut {NoStop}%
\bibitem [{\citenamefont {\ifmmode \mbox{\c{C}}\else
  \c{C}\fi{}ift\ifmmode~\mbox{\c{c}}\else \c{c}\fi{}i}\ and\ \citenamefont
  {Waalkens}(2013)}]{Waalkens13}%
  \BibitemOpen
  \bibfield  {author} {\bibinfo {author} {\bibfnamefont {U.}~\bibnamefont
  {\ifmmode \mbox{\c{C}}\else \c{C}\fi{}ift\ifmmode~\mbox{\c{c}}\else
  \c{c}\fi{}i}}\ and\ \bibinfo {author} {\bibfnamefont {H.}~\bibnamefont
  {Waalkens}},\ }\href@noop {} {\bibfield  {journal} {\bibinfo  {journal}
  {Phys. Rev. Lett.}\ }\textbf {\bibinfo {volume} {110}},\ \bibinfo {pages}
  {233201} (\bibinfo {year} {2013})}\BibitemShut {NoStop}%
\bibitem [{\citenamefont {Bardakcioglu}\ \emph {et~al.}(2018)\citenamefont
  {Bardakcioglu}, \citenamefont {Junginger}, \citenamefont {Feldmaier},
  \citenamefont {Main},\ and\ \citenamefont {Hernandez}}]{hern18g}%
  \BibitemOpen
  \bibfield  {author} {\bibinfo {author} {\bibfnamefont {R.}~\bibnamefont
  {Bardakcioglu}}, \bibinfo {author} {\bibfnamefont {A.}~\bibnamefont
  {Junginger}}, \bibinfo {author} {\bibfnamefont {M.}~\bibnamefont
  {Feldmaier}}, \bibinfo {author} {\bibfnamefont {J.}~\bibnamefont {Main}}, \
  and\ \bibinfo {author} {\bibfnamefont {R.}~\bibnamefont {Hernandez}},\
  }\href@noop {} {\bibfield  {journal} {\bibinfo  {journal} {Phys. Rev. E}\
  }\textbf {\bibinfo {volume} {98}},\ \bibinfo {pages} {032204} (\bibinfo
  {year} {2018})}\BibitemShut {NoStop}%
\bibitem [{\citenamefont {Schraft}\ \emph {et~al.}(2018)\citenamefont
  {Schraft}, \citenamefont {Junginger}, \citenamefont {Feldmaier},
  \citenamefont {Bardakcioglu}, \citenamefont {Main}, \citenamefont {Wunner},\
  and\ \citenamefont {Hernandez}}]{hern18c}%
  \BibitemOpen
  \bibfield  {author} {\bibinfo {author} {\bibfnamefont {P.}~\bibnamefont
  {Schraft}}, \bibinfo {author} {\bibfnamefont {A.}~\bibnamefont {Junginger}},
  \bibinfo {author} {\bibfnamefont {M.}~\bibnamefont {Feldmaier}}, \bibinfo
  {author} {\bibfnamefont {R.}~\bibnamefont {Bardakcioglu}}, \bibinfo {author}
  {\bibfnamefont {J.}~\bibnamefont {Main}}, \bibinfo {author} {\bibfnamefont
  {G.}~\bibnamefont {Wunner}}, \ and\ \bibinfo {author} {\bibfnamefont
  {R.}~\bibnamefont {Hernandez}},\ }\href@noop {} {\bibfield  {journal}
  {\bibinfo  {journal} {Phys. Rev. E}\ }\textbf {\bibinfo {volume} {97}},\
  \bibinfo {pages} {042309} (\bibinfo {year} {2018})}\BibitemShut {NoStop}%
\bibitem [{\citenamefont {Craven}, \citenamefont {Bartsch},\ and\ \citenamefont
  {Hernandez}(2014)}]{hern14f}%
  \BibitemOpen
  \bibfield  {author} {\bibinfo {author} {\bibfnamefont {G.~T.}\ \bibnamefont
  {Craven}}, \bibinfo {author} {\bibfnamefont {T.}~\bibnamefont {Bartsch}}, \
  and\ \bibinfo {author} {\bibfnamefont {R.}~\bibnamefont {Hernandez}},\
  }\href@noop {} {\bibfield  {journal} {\bibinfo  {journal} {J. Chem. Phys.}\
  }\textbf {\bibinfo {volume} {141}},\ \bibinfo {pages} {041106} (\bibinfo
  {year} {2014})}\BibitemShut {NoStop}%
\bibitem [{\citenamefont {Tsch{\"o}pe}\ \emph {et~al.}()\citenamefont
  {Tsch{\"o}pe}, \citenamefont {Feldmaier}, \citenamefont {Main},\ and\
  \citenamefont {Hernandez}}]{hern19T1}%
  \BibitemOpen
  \bibfield  {author} {\bibinfo {author} {\bibfnamefont {M.}~\bibnamefont
  {Tsch{\"o}pe}}, \bibinfo {author} {\bibfnamefont {M.}~\bibnamefont
  {Feldmaier}}, \bibinfo {author} {\bibfnamefont {J.}~\bibnamefont {Main}}, \
  and\ \bibinfo {author} {\bibfnamefont {R.}~\bibnamefont {Hernandez}},\
  }\href@noop {} {}\bibinfo {note} {``Neural network approach for the dynamics
  on the normally hyperbolic invariant manifold of periodically driven
  systems,'' {\it Phys. Rev. E}, submitted}\BibitemShut {NoStop}%
\bibitem [{\citenamefont {Townsend}\ \emph {et~al.}(2004)\citenamefont
  {Townsend}, \citenamefont {Lahankar}, \citenamefont {Lee}, \citenamefont
  {Chambreau}, \citenamefont {Suits}, \citenamefont {Zhang}, \citenamefont
  {Rheinecker}, \citenamefont {Harding},\ and\ \citenamefont
  {Bowman}}]{bowman04a}%
  \BibitemOpen
  \bibfield  {author} {\bibinfo {author} {\bibfnamefont {D.}~\bibnamefont
  {Townsend}}, \bibinfo {author} {\bibfnamefont {S.~A.}\ \bibnamefont
  {Lahankar}}, \bibinfo {author} {\bibfnamefont {S.~K.}\ \bibnamefont {Lee}},
  \bibinfo {author} {\bibfnamefont {S.~D.}\ \bibnamefont {Chambreau}}, \bibinfo
  {author} {\bibfnamefont {A.~G.}\ \bibnamefont {Suits}}, \bibinfo {author}
  {\bibfnamefont {X.}~\bibnamefont {Zhang}}, \bibinfo {author} {\bibfnamefont
  {J.~L.}\ \bibnamefont {Rheinecker}}, \bibinfo {author} {\bibfnamefont
  {L.~B.}\ \bibnamefont {Harding}}, \ and\ \bibinfo {author} {\bibfnamefont
  {J.~M.}\ \bibnamefont {Bowman}},\ }\href@noop {} {\bibfield  {journal}
  {\bibinfo  {journal} {Science}\ }\textbf {\bibinfo {volume} {306}},\ \bibinfo
  {pages} {1158} (\bibinfo {year} {2004})}\BibitemShut {NoStop}%
\bibitem [{\citenamefont {Ulusoy}, \citenamefont {Stanton},\ and\ \citenamefont
  {Hernandez}(2013)}]{hern13e}%
  \BibitemOpen
  \bibfield  {author} {\bibinfo {author} {\bibfnamefont {I.~S.}\ \bibnamefont
  {Ulusoy}}, \bibinfo {author} {\bibfnamefont {J.~F.}\ \bibnamefont {Stanton}},
  \ and\ \bibinfo {author} {\bibfnamefont {R.}~\bibnamefont {Hernandez}},\
  }\href@noop {} {\bibfield  {journal} {\bibinfo  {journal} {J. Phys. Chem. A}\
  }\textbf {\bibinfo {volume} {117}},\ \bibinfo {pages} {10567} (\bibinfo
  {year} {2013})}\BibitemShut {NoStop}%
\bibitem [{\citenamefont {Maugi\`{e}re}\ \emph {et~al.}(2014)\citenamefont
  {Maugi\`{e}re}, \citenamefont {Collins}, \citenamefont {Ezra}, \citenamefont
  {Farantos},\ and\ \citenamefont {Wiggins}}]{wiggins14a}%
  \BibitemOpen
  \bibfield  {author} {\bibinfo {author} {\bibfnamefont {F.~A.~L.}\
  \bibnamefont {Maugi\`{e}re}}, \bibinfo {author} {\bibfnamefont
  {P.}~\bibnamefont {Collins}}, \bibinfo {author} {\bibfnamefont
  {G.}~\bibnamefont {Ezra}}, \bibinfo {author} {\bibfnamefont {S.~C.}\
  \bibnamefont {Farantos}}, \ and\ \bibinfo {author} {\bibfnamefont
  {S.}~\bibnamefont {Wiggins}},\ }\href@noop {} {\bibfield  {journal} {\bibinfo
   {journal} {Chem. Phys. Lett.}\ }\textbf {\bibinfo {volume} {592}},\ \bibinfo
  {pages} {282} (\bibinfo {year} {2014})}\BibitemShut {NoStop}%
\bibitem [{\citenamefont {Bowman}\ and\ \citenamefont
  {Suits}(2011)}]{bowman2011c}%
  \BibitemOpen
  \bibfield  {author} {\bibinfo {author} {\bibfnamefont {J.~M.}\ \bibnamefont
  {Bowman}}\ and\ \bibinfo {author} {\bibfnamefont {A.~G.}\ \bibnamefont
  {Suits}},\ }\href@noop {} {\bibfield  {journal} {\bibinfo  {journal} {Phys.
  Today}\ }\textbf {\bibinfo {volume} {64}},\ \bibinfo {pages} {33} (\bibinfo
  {year} {2011})}\BibitemShut {NoStop}%
\bibitem [{\citenamefont {Bowman}(2014)}]{bowman14a}%
  \BibitemOpen
  \bibfield  {author} {\bibinfo {author} {\bibfnamefont {J.~M.}\ \bibnamefont
  {Bowman}},\ }\href@noop {} {\bibfield  {journal} {\bibinfo  {journal} {Mol.
  Phys.}\ }\textbf {\bibinfo {volume} {112}},\ \bibinfo {pages} {2516}
  (\bibinfo {year} {2014})}\BibitemShut {NoStop}%
\bibitem [{\citenamefont {H{\"a}nggi}, \citenamefont {Talkner},\ and\
  \citenamefont {Borkovec}(1990)}]{rmp90}%
  \BibitemOpen
  \bibfield  {author} {\bibinfo {author} {\bibfnamefont {P.}~\bibnamefont
  {H{\"a}nggi}}, \bibinfo {author} {\bibfnamefont {P.}~\bibnamefont {Talkner}},
  \ and\ \bibinfo {author} {\bibfnamefont {M.}~\bibnamefont {Borkovec}},\
  }\href {http://link.aps.org/doi/10.1103/RevModPhys.62.251} {\bibfield
  {journal} {\bibinfo  {journal} {Rev. Mod. Phys.}\ }\textbf {\bibinfo {volume}
  {62}},\ \bibinfo {pages} {251} (\bibinfo {year} {1990})},\ \bibinfo {note}
  {and references therein}\BibitemShut {NoStop}%
\bibitem [{\citenamefont {Craven}, \citenamefont {Bartsch},\ and\ \citenamefont
  {Hernandez}(2015)}]{hern15a}%
  \BibitemOpen
  \bibfield  {author} {\bibinfo {author} {\bibfnamefont {G.~T.}\ \bibnamefont
  {Craven}}, \bibinfo {author} {\bibfnamefont {T.}~\bibnamefont {Bartsch}}, \
  and\ \bibinfo {author} {\bibfnamefont {R.}~\bibnamefont {Hernandez}},\
  }\href@noop {} {\bibfield  {journal} {\bibinfo  {journal} {J. Chem. Phys.}\
  }\textbf {\bibinfo {volume} {142}},\ \bibinfo {pages} {074108} (\bibinfo
  {year} {2015})}\BibitemShut {NoStop}%
\bibitem [{\citenamefont {Upadhyay}(2006)}]{upadhyay06}%
  \BibitemOpen
  \bibfield  {author} {\bibinfo {author} {\bibfnamefont {S.~K.}\ \bibnamefont
  {Upadhyay}},\ }\href@noop {} {\emph {\bibinfo {title} {Chemical Kinetics and
  Reaction Dynamics}}}\ (\bibinfo  {publisher} {Springer Netherlands},\
  \bibinfo {year} {2006})\BibitemShut {NoStop}%
\bibitem [{\citenamefont {Connors}(1990)}]{connors90}%
  \BibitemOpen
  \bibfield  {author} {\bibinfo {author} {\bibfnamefont {K.~A.}\ \bibnamefont
  {Connors}},\ }\href@noop {} {\emph {\bibinfo {title} {Chemical Kinetics: The
  Study of Reaction Rates in Solution}}}\ (\bibinfo  {publisher}
  {Wiley-{VCH}},\ \bibinfo {year} {1990})\BibitemShut {NoStop}%
\bibitem [{\citenamefont {{P. H{\"a}nggi}}, \citenamefont {{P. Talkner}},\ and\
  \citenamefont {{M. Borkovec}}(1990)}]{Haenggi1990}%
  \BibitemOpen
  \bibfield  {author} {\bibinfo {author} {\bibnamefont {{P. H{\"a}nggi}}},
  \bibinfo {author} {\bibnamefont {{P. Talkner}}}, \ and\ \bibinfo {author}
  {\bibnamefont {{M. Borkovec}}},\ }\href@noop {} {\bibfield  {journal}
  {\bibinfo  {journal} {Rev. Mod. Phys.}\ }\textbf {\bibinfo {volume} {62}},\
  \bibinfo {pages} {251} (\bibinfo {year} {1990})}\BibitemShut {NoStop}%
\bibitem [{\citenamefont {Revuelta}\ \emph {et~al.}(2017)\citenamefont
  {Revuelta}, \citenamefont {Craven}, \citenamefont {Bartsch}, \citenamefont
  {Borondo}, \citenamefont {Benito},\ and\ \citenamefont
  {Hernandez}}]{hern17f}%
  \BibitemOpen
  \bibfield  {author} {\bibinfo {author} {\bibfnamefont {F.}~\bibnamefont
  {Revuelta}}, \bibinfo {author} {\bibfnamefont {G.~T.}\ \bibnamefont
  {Craven}}, \bibinfo {author} {\bibfnamefont {T.}~\bibnamefont {Bartsch}},
  \bibinfo {author} {\bibfnamefont {F.}~\bibnamefont {Borondo}}, \bibinfo
  {author} {\bibfnamefont {R.~M.}\ \bibnamefont {Benito}}, \ and\ \bibinfo
  {author} {\bibfnamefont {R.}~\bibnamefont {Hernandez}},\ }\href@noop {}
  {\bibfield  {journal} {\bibinfo  {journal} {J. Chem. Phys.}\ }\textbf
  {\bibinfo {volume} {147}},\ \bibinfo {pages} {074104} (\bibinfo {year}
  {2017})}\BibitemShut {NoStop}%
\bibitem [{\citenamefont {Junginger}\ \emph {et~al.}(2017)\citenamefont
  {Junginger}, \citenamefont {Duvenbeck}, \citenamefont {Feldmaier},
  \citenamefont {Main}, \citenamefont {Wunner},\ and\ \citenamefont
  {Hernandez}}]{hern17e}%
  \BibitemOpen
  \bibfield  {author} {\bibinfo {author} {\bibfnamefont {A.}~\bibnamefont
  {Junginger}}, \bibinfo {author} {\bibfnamefont {L.}~\bibnamefont
  {Duvenbeck}}, \bibinfo {author} {\bibfnamefont {M.}~\bibnamefont
  {Feldmaier}}, \bibinfo {author} {\bibfnamefont {J.}~\bibnamefont {Main}},
  \bibinfo {author} {\bibfnamefont {G.}~\bibnamefont {Wunner}}, \ and\ \bibinfo
  {author} {\bibfnamefont {R.}~\bibnamefont {Hernandez}},\ }\href@noop {}
  {\bibfield  {journal} {\bibinfo  {journal} {J. Chem. Phys.}\ }\textbf
  {\bibinfo {volume} {147}},\ \bibinfo {pages} {06401} (\bibinfo {year}
  {2017})}\BibitemShut {NoStop}%
\bibitem [{\citenamefont {{P\'arraga}}\ \emph {et~al.}(2018)\citenamefont
  {{P\'arraga}}, \citenamefont {Arranz}, \citenamefont {Benito},\ and\
  \citenamefont {Borondo}}]{borondo18a}%
  \BibitemOpen
  \bibfield  {author} {\bibinfo {author} {\bibfnamefont {H.}~\bibnamefont
  {{P\'arraga}}}, \bibinfo {author} {\bibfnamefont {F.~J.}\ \bibnamefont
  {Arranz}}, \bibinfo {author} {\bibfnamefont {R.~M.}\ \bibnamefont {Benito}},
  \ and\ \bibinfo {author} {\bibfnamefont {F.}~\bibnamefont {Borondo}},\
  }\href@noop {} {\bibfield  {journal} {\bibinfo  {journal} {J. Phys. Chem. A}\
  }\textbf {\bibinfo {volume} {122}},\ \bibinfo {pages} {3433} (\bibinfo {year}
  {2018})}\BibitemShut {NoStop}%
\end{thebibliography}%

\end{document}